\documentclass[%
preprint,
nofootinbib,
 amsmath,amssymb,
 aps,
prd,
]{revtex4-2}

\usepackage[colorlinks,linkcolor=refcol,citecolor=refcol,urlcolor=refcol]{hyperref}
\usepackage{natbib}

\usepackage{tikz-feynman}
\usepackage{graphicx}
\usepackage{bm}

\usepackage{color}
\usepackage[makeroom]{cancel}
\definecolor{refcol}{rgb}{0,0,.5}
\usepackage{microtype}

\definecolor{red}{rgb}{1,0,0}
\definecolor{blue}{rgb}{0,0,1}

\usepackage{amsmath}

\newcommand*{\scv}{z}

\begin{document}

\title{Universal location of Yang-Lee edge singularity in classic O(N) universality classes }

\author{Gregory Johnson}
\affiliation{Department of Physics, North Carolina State University, Raleigh, NC 27695, USA}

\author{Fabian Rennecke}
 \email{fabian.rennecke@theo.physik.uni-giessen.de} 
    \affiliation{Institute for Theoretical Physics, Justus Liebig University Giessen, Heinrich-Buff-Ring 16, 35392 Giessen, Germany}
   \affiliation{Helmholtz Research Academy Hesse for FAIR (HFHF), Campus Giessen, 35392 Giessen, Germany}

\author{Vladimir V. Skokov}
\email{VSkokov@ncsu.edu} 
    \affiliation{Department of Physics, North Carolina State University, Raleigh, NC 27695, USA}
    \affiliation{RIKEN BNL Research Center, Brookhaven National Laboratory, Upton, NY 11973, USA}

\begin{abstract}
Employing the functional renormalization group approach at next-to-leading order of  the derivative expansion, we refine our earlier findings for the location of the Yang-Lee edge singularity in classic O(N) universality classes.  
For the universality classes of interest to QCD, in three dimensions, we found   $|\scv_c|/R_\chi^{1/\gamma} = 1.612(9),  1.597(3)$ for $N=2$, $4$ correspondingly.  We also established  $|\scv_c| = 2.04(8),  1.69(3)$ for $N=2$, $4$ albeit with greater systematic error.
\end{abstract}

\date{\today}

\maketitle


\newpage 

\tableofcontents

\section{Introduction}
\label{sec:intro}
Due to the divergence of the correlation length near a second-order phase transition, the dynamics of the system becomes independent of the microscopic details and only reflects the grand properties -- the dimensionality and global symmetries. This allows one to collect systems of varied microscopic origin into a limited number of universality classes. By studying one member of such a class,  the emergent universal behavior allows one to establish the properties of many different systems regardless of their microscopic complexity. 

Thus, it is not surprising that for the most ubiquitous classic  universality classes of $O(N)$ systems many universal properties (such as critical exponents, critical universal amplitudes, critical equations of state)  are known with extreme precision, see e.g. Refs.~\cite{Campostrini:2002ky,Engels:2002fi,Hasenbusch:2008zz,Kos:2013tga,Hasenbusch:2021rse} and references therein. One notable exception is the universal location of the Yang-Lee edge singularity, which  was only recently determined in Refs.~\cite{Connelly:2020gwa,Rennecke:2022ohx}. In this paper, we continue to refine these results.  

\begin{figure}
    \centering
       \includegraphics[width=0.49\linewidth]{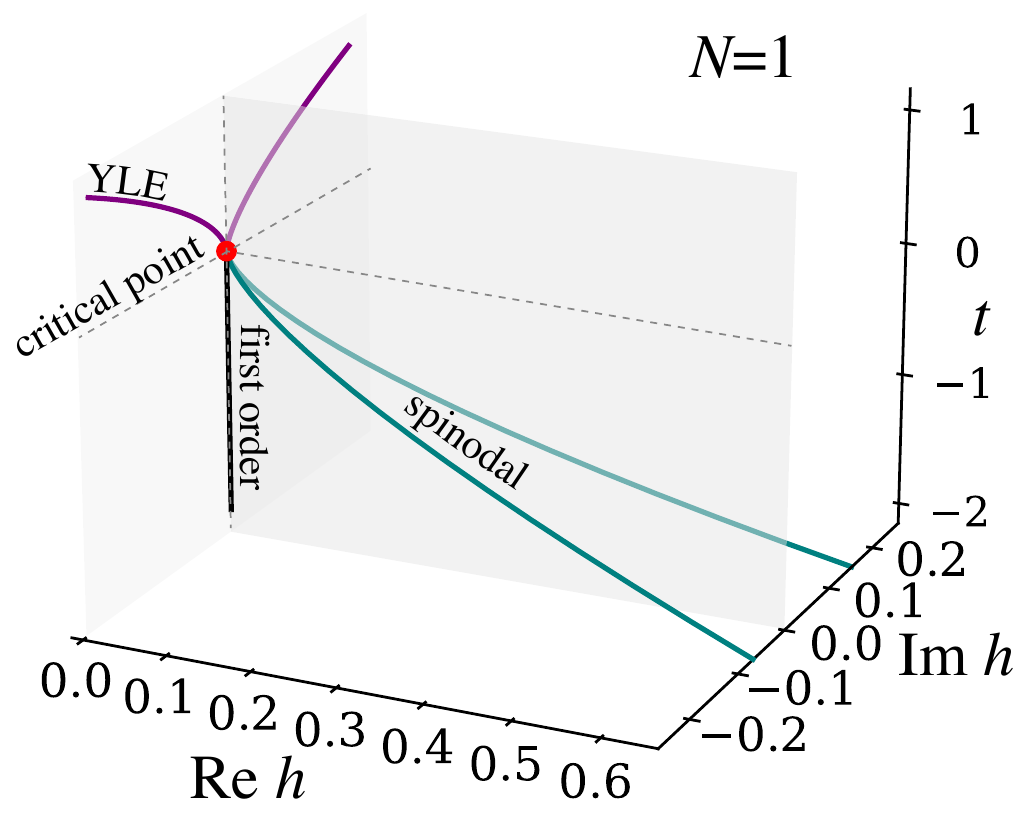}
    \caption{Analytic structure of the universal phase diagram for $N=1$. Only branch points are displayed; the cuts are omitted for the clarity of the figure. To draw this figure, we used $h = (t/\scv_c)^{\beta \delta}$ with the realistic critical exponents and the location of the singularity obtained $|\scv_c|$ from    Ref.~\cite{Rennecke:2022ohx} complemented by the value of  $R_\chi = 1.72$ from Ref.~\cite{Engels:2002fi}.  For the mean-field equation of state and for the large $N$ limit in $d=3$, the spinodals are located on the real $h$  axis due to integer values of  $2\beta\delta$.  Only positive values for real values of $h$ are shown.     }
    \label{fig:illustra}
\end{figure}

Lee and Yang  demonstrated an intimate connection between the analytical structure of the equation of state and the phase structure~\cite{PhysRev.87.404,PhysRev.87.410}. Specifically, in the symmetric phase, the Lee-Yang theorem states that the equations of state of O$(N)$-symmetric $\phi^4$ theories have a branch cut at purely imaginary values of the magnetic field $h$. The cut terminates at two branch points -- the Yang-Lee edge singularities. 
A second-order (first-order) phase transition at $t\propto T-T_c = 0$ occurs when the singularities pinch (cross) the real $h$-axis.
In the broken phase, the singularities are also known as spinodals, see Fig.~\ref{fig:illustra}. 
Remarkably, the edge singularities can be seen as  critical points themselves. As illustrated in Fig.~\ref{fig:illustra}, variation of only one parameter $h$ allows to tune the system to these critical points.  In contrast, the conventional Wilson-Fisher critical point requires tuning two parameters $t$ and $h$. This not only signifies the greater ontological importance of the Yang-Lee edge (YLE) singularity but also determines the number of independent critical exponents. At the Wilson-Fisher critical point, there are two relevant perturbations and thus two independent critical exponents. At the YLE, it follows that there is only one independent critical exponent $\sigma_{\rm YLE} = 1/\delta_{\rm YLE}$. It determines the scaling of the magnetization, $M\sim M_{\rm c} + (h- h_c)^{\sigma_{\rm YLE}}$, where $M_c$ and $h_c$ are purely imaginary. The numerical value of the edge critical exponent in three dimensions (and for any $N$ of the underlying universality class) has been determined by a variety of methods, see e.g. Refs.~\cite{Gliozzi:2014jsa, An:2016lni, Zambelli:2016cbw,Borinsky:2021jdb}. 

An interesting property of the YLE critical point is that it is characterized by a $\phi^3$ theory and consequently its upper critical dimension is  6 \cite{Fisher:1978pf}.
Therefore, the conventional $\varepsilon$ expansion near four dimensions
applied to study the Wilson-Fisher critical point of the  underlying universality class has only limited predictive power for locating the YLE singularity. We come back to this in more detail in Sect.~\ref{Sec:Epsilon}.

The numerical calculations in this paper are performed using the Functional Renormalization Group (FRG) approach, see Ref.~\cite{Berges:2000ew} for a review. We extend the results  of our previous work (see Refs.~\cite{Connelly:2020gwa} and \cite{Rennecke:2022ohx}) significantly. First, we improve the truncation scheme going to the 
 {(truncated)} first order derivative expansion and including the dependence of the wave function  renormalizations on the field for  $N>1$. In our original study~\cite{Connelly:2020gwa}, the calculations were performed in the so-called LPA' approximation which assumes  a field-independent wave function renormalization, while in Ref.~\cite{Rennecke:2022ohx} we only investigated the Ising universality class, $N=1$. Second, we accounted for the residual dependence on the regulator by performing a minimal sensitivity analysis~\cite{Canet:2002gs} which was motivated by 
minimizing the sensitivity to nonphysical parameters in conventional perturbation theory with different renormalization schemes~\cite{Stevenson:1981vj}.   

The paper is organized as follows. We start by defining a required set of universal quantities, functions and non-universal metric factors in Sect.~\ref{Sec:Scaling}. We then review analytical results for the location of the YLE singularity in Sect.~\ref{Sec:Analytic}: the large $N$ limit and for the number of spatial dimensions close to 4. For the number of components $N\ne 1$, we discuss the behavior of the singularity near two dimensions.   In Section \ref{Sec:FRG}, we turn to  FRG calculations where we extract the location of the singularity for various $N$ in three spatial dimensions.  We end with conclusions in Sect.~\ref{Sec:Concl}.

\section{Scaling equation, critical amplitudes and  exponents}
\label{Sec:Scaling}
Consider a system near a critical point with two relevant parameters $t$ and $h$ introduced in a such a way as to detune the system from criticality which occurs at $t=h=0$. We will refer to $t$ as the temperature. Its defining property is that non-zero values of $t$ do not explicitly break any symmetries of the system. However, a non-zero $t$ may lead to a spontaneous symmetry breaking either for positive or negative $t$. Conventionally we assign positive values of $t$ to when the spontaneous symmetry breaking is not possible -- in other words, $t>0$ defines the symmetric phase of the system.   In contrast to $t$, non-zero values of $h$, to which we will refer to as the external magnetic field, break the symmetry explicitly. We quantify the system's response to $t$ and $h$ by measuring the order parameter, to which we also will refer to as magnetization $M$.     

The renormalization group analysis (see e.g. Ref.~\cite{Amit:1984ms})  demonstrates 
that the equations of state describing the dependence of the magnetization on the parameters $t$ and $h$  has a homogeneous form and can be written as 
\begin{align}
      h = M^\delta f (x\equiv t M^{-1/\beta})\,,
\end{align}
where $\beta$ and $\delta$ are universal critical exponents, and $f(x)$ is a universal scaling function.  

In general, the parameters $t$ and $h$ are related to the physical parameters of the system through two non-universal proportionality coefficients, also called metric factors. The metric factors are usually chosen in a such a way as to satisfy two normalization conditions for the function $f$: 
\begin{align}
    &f(0) = 1 , \\
    &f(-1) = 0. 
\end{align}
The above form of the scaling equation of state was suggested by Widom~\cite{doi:10.1063/1.1696618}. One of its advantages is that it can be straightforwardly derived using the $\varepsilon$ expansion. Its disadvantage is that it leads to an implicit dependence of $M$ on $t$ and $h$. An alternative form 
\begin{equation}\label{eq:fG}
   M(t,h) = h^{1/\delta} f_G( \scv = t\, / h^{1/\Delta})
\end{equation}
solves this issue. Here we have introduced the so-called gap critical exponent, $\Delta=\beta\delta$. 
The function $f_G$ is a function of one variable; it  encodes most of the critical statics. It has to satisfy the normalization conditions 
\begin{align}
\label{Eq:Normalization}
        f_G(0)  &=  1, \\
        \lim_{\scv\to - \infty} f_G(\scv) &\to (-\scv)^\beta
\end{align}
to be consistent with the Widom scaling.  
As we alluded to before, the set of the normalization conditions requires the redefinition of the non-universal parameters $t$ and $h$. 
Generically near a critical point we have 
    \begin{align}
        M &= B_c {\hat h}^{1/\delta}, \quad \hat t=0, \\
        M &= B (-\hat t)^\beta, \quad \hat H=0 \quad \text{and} \quad  \hat t<0\,.  
    \end{align}
The normalization conditions require us to define $t$ and $h$ in a such a way as to absorb the prefactors $B$ and $B_c$: 
    \begin{align}
        M &=  {h}^{1/\delta}, \quad t=0, \\
        M &= (-t)^\beta, \quad h=0 \quad \text{and} \quad   t<0\,.  
    \end{align}
    
In Sect.~\ref{Sec:FRG}, we formulate the FRG approach to locating the YLE singularity. As it  is our primary objective, our truncation method is optimized to perform simulations in the symmetric phase. Calculations in the broken phase are possible in a different truncation scheme; however, we want to extract all required quantities within one scheme to avoid introducing systematic errors by mixing different truncations in the simulations. We thus strive to avoid the broken phase. This motivates us to introduce another universal quantity 
\begin{align}
    \label{Eq:zeta}
    \zeta = \frac{\scv}{R_\chi^{1/\gamma}}\,
\end{align}
where the universal ratio $R_\chi$ is defined by the limit 
\begin{align}
\label{Eq:Rchidef}
R_\chi  = \lim_{\scv\to \infty} f_G(\scv) \scv^\gamma
\end{align}
and $\gamma$ is the critical exponent connected to $\delta$ and $\beta$ through the scaling relation: 
\begin{align}\label{Eq:nusc}
     \gamma = \Delta - \beta\,. 
\end{align}
The asymptotic behaviour of the function $f_G(\scv)$ at large argument has a simple physical origin: the magnetization has to be a linear function of $h$ in the symmetric phase $t>0$. From Eq.~\eqref{eq:fG} follows that the scaling function $f_G(z)$ therefore has to go like $z^{-(\Delta -\beta)} = z^{-\gamma}$ leading to  the identity Eq.~\eqref{Eq:nusc}. 
Note that working in the symmetric phase allows us to directly extract the critical exponent $\gamma$ through the scaling for the magnetic susceptibility 
    \begin{align}
        \label{Eq:Chi}
        \chi(\hat t,\hat h=0) = \frac{ \partial M} {\partial \hat h}  = C_+ \hat{t} ^{-\gamma}\,.
    \end{align}
Using this expression it is straightforward to show that 
\begin{align}
    R_\chi = \frac{C^+ B^{\delta -1 }}{B_c^\delta}\,
\end{align}
and that the introduced $\zeta$ is independent of the amplitude $B$:
\begin{align}
\label{Eq:zetaAmpl}
    \zeta = \left( \frac{ B_c} { C_+ }  \right)^{1/\gamma} \frac{ \hat t  } { {\hat H}^{1/\Delta} } \, 
\end{align}
thus explicitly demonstrating that in order to extract the location of the YLE singularity in $\zeta$ we 
do not need to perform simulations in the broken phase. 

We stress  that $\zeta$ and $z$ are related through a universal number $R_\chi$ and universal critical exponent $\gamma$.  On one hand, $R_\chi$ can be computed in the FRG approach\footnote{Precision calculations were performed in a state of the art study in Ref.~\cite{DePolsi:2021cmi}.} but would require probing the broken phase and thus switching the FRG truncation scheme used in this paper. The associated systematic error is difficult to assess. On the other hand, for applications to lattice QCD, $|z_c|$ is often considered. We thus will provide a separate set of results for $|z_c|$ using $R_\chi$  obtained in Ref.~\cite{DePolsi:2021cmi}.

Finally, for the purpose of the next sections,  we also introduce the anomalous dimension critical exponent $\eta$. It describes the power law dependence of the static correlation function on the distance  at the critical point. In $d$ spatial dimensions: 
\begin{align}
    G(|x|) \sim |x| ^ {-(d-2+\eta)}\,.  
\end{align}
The anomalous dimension satisfies the following scaling relation~\cite{Amit:1984ms}   
\begin{align}    \label{Eq:etasc}
    2 - \eta &= d\, \frac{ \delta -1}{\delta+1}\,.
\end{align}

\section{Analytical results for  location of Yang-Lee Edge singularity} 
\label{Sec:Analytic}
We remind the reader that the Yang-Lee edge singularities are branch points of the function $f_G(\scv)$ in the symmetric phase $t>0$. They can be determined by finding zeros of the inverse magnetic field susceptibility. Most generally, the Lee-Yang theorem~\cite{PhysRev.87.404,PhysRev.87.410} 
implies that they have to be located on the imaginary $h$ axis. Thus the argument of the singularity $z_c$  (and its complex conjugate) is fully determined by the critical exponents of the underlying  O($N$) universality class:
\begin{align}
    \scv_c = |\scv_c| e^{i \frac{\pi}{2\Delta}}\,.
\end{align}
Note that the argument of $\zeta_c$ coincides with that of $\scv_c$.

As far as the absolute value of the location of the YLE singularity $|\scv_c|$ is concerned, O($N$) universality classes with $N>1$  do not enjoy many (neither exact nor approximate) analytical results. Notable exceptions are $N\to\infty$ limit and the theory near four dimensions. We detail both analytical results below.

\subsection{ $N\to \infty$ limit} 
\label{Sec:LargeN}
The large N scaling equation of state can be readily computed; for a review, see Ref.~\cite{Moshe:2003xn}, where the Widom scaling relation and critical exponents  were derived: 
\begin{align}
f = (1+x)^{2/(d-2)}\,
\end{align}
and 
\begin{align}
    \delta &= \frac{d+2}{d-2},\quad  \beta = \frac{1}{2}\,.
\end{align}
The function $f$ defines the value of  $\gamma = 2/(d-2)$ at asymptotically large values of $x$
and $R_\chi = 1$ following Eq.~\eqref{Eq:Rchidef}.  To find the position of the YLE singularity, it is convenient to consider the inverse magnetic field susceptibility $\chi^{-1} = \left. \frac{\partial h}{\partial M}\right|_t$. Its zeroes define the values of $x$ at the YLE, $x_c$. In terms of the function $f(x)$, we have 
\begin{align}
\label{Eq:Fcritic}
    \beta \delta f(x_c) - x_c f'(x_c) = 0\,.
\end{align} 
This leads to 
\begin{align}
    x_c &= -\frac{d+2}{d-2}\,, \\ 
    f_c &= \left(\frac{4}{2-d}\right)^{\frac{2}{d-2}}\,.
\end{align}
Now we can proceed with finding $\scv_c$. For this we express $z_c$ in terms of $x$ and $f$: 
\begin{align}
    \scv_c = \frac { t_c } {h_c^{1/\beta \delta}}  = \frac{ x_c} { f_c^{1/\beta \delta} } \,
\end{align}
leading to  
\begin{align}
    |\scv_c| = \frac{d+2} {2 ^{\frac{8}{d+2}}} (d-2) ^ {\frac{2-d}{2+d}}\,.
\end{align}
At large $N$, $R_\chi = 1$, thus    $|\zeta_c| = |\scv_c| $. 
To compare to the result of the next section, we perform the expansion near four dimensions $d=4-\varepsilon$: 
\begin{align}
   |\scv_c| \approx  |\scv_c^{\rm MF}|  \left(1 - \frac{\ln 2}{9} \varepsilon \right) \,
    \label{Eq:LargeN_near4}
\end{align}
where we introduced 
$\scv_c^{\rm MF}$ as the value of $\scv_c$ at the upper critical dimension of the O($N$) universality classes  $d=4$ and the corresponding 
absolute value $|\scv_c^{\rm MF}| = \frac{3}{2^{2/3}} $. 

Near the lower critical dimension  $d=2+ \tilde \varepsilon$, we obtain
\begin{align}
   |\scv_c| \approx 1 + \frac{1}{4} \left(1 - \ln \frac{ \tilde \varepsilon}{4}  \right)  \tilde \varepsilon \,.
    \label{Eq:LargeN_near2}
\end{align}

\subsection{The $\varepsilon$-expansion} 
\label{Sec:Epsilon}
Although $N\to \infty$ limit provides an analytic result for any $d$ in the range $2<d<4$, it is not well suited to describe phenomenologically relevant universality classes. Specifically for finite temperature QCD we are interested in the Ising universality class  $N=1$ and the Heisenberg model with $N=2$ (due to lattice discretization artifacts, see e.g. Ref.~\cite{HotQCD:2019xnw})  and $N=4$. There is another analytic limit in which one can perform the calculation -- near the upper critical dimension $d=4-\varepsilon$. As we document below, as far as the position of the YLE singularity is concerned, the utility of this approach is somewhat restricted and it  cannot be systematically improved to yield a reliable result in three dimensions. However, it provides some useful information on the location of the YLE singularity  near four dimensions. It is also not limited to $N\to \infty$. Moreover, it serves an estimate of the value of $N$ at which one can apply a large N approximation for the  purpose of locating the YLE singularity.         

In the conventional $\varepsilon$ expansion, see e.g. Ref.~\cite{Amit:1984ms}, near the upper critical dimension $d=4-\varepsilon$, the critical exponents are   
\begin{align}
    \gamma&=1+\frac{N+2}{2(N+8)}\varepsilon
+O\left(\varepsilon^2\right)\,,  \\ 
\beta \delta &=
\frac{3}{2} + \frac{1}{2} \left(1 - \frac{9}{N + 8}\right) \varepsilon  +O\left(\varepsilon^2\right)\,.
\end{align}
The same method yields  the universal amplitude ratio, see Ref.~\cite{Abe:1978aq}, 
\begin{align}
\label{Eq:Rchi}
    R_\chi = 
    1 + \frac{3}{2 (N + 8)} \ln \left(\frac{27}{4}\right) \varepsilon  + O\left(\varepsilon^2\right)\,.
\end{align}
Note that, in the large $N$ limit, this  is consistent with  the previous subsection, $R_\chi=1$. 

To the linear order in $\varepsilon$~\cite{1972JETPL..16..178A}, the scaling function $f$ has the following form 
\begin{align}
    f(x) &= 1 + x  \notag \\ 
    & + \varepsilon \frac{
      (N-1) (x+1) \ln (x+1)+ 
      9(x+3) \ln (x+3)
      - 9 \ln 3 + 
      3 x \ln \frac{4}{27} 
    }{2(N+8)}  + {\cal O} (\varepsilon^2)\,.
\end{align}

We now turn to the location of the YLE singularity. At the leading order $\varepsilon$-expansion, we  obtain $x^{(0)}_c = -3$  for the solution of Eq.~\eqref{Eq:Fcritic}.
For our purpose, 
the exact expression for $\hat{x}(\varepsilon)$  is of no importance as will be demonstrated below. 
Moreover the first correction  $\hat{x}(\varepsilon)$ to the leading order $x_c^{(0)}$ is already non-perturbative {(See Appendix~\ref{sec:appendixnonp} for details).}   

To the first order in $\varepsilon$, 
\begin{align}
f_c = f(x_c) = 
-2 + \frac{\varepsilon}{2 (N + 8)} \left[-2 (N - 1) \ln(-2) + 18 \ln \left(\frac{3}{2}\right) \right]+  \hat{x}(\varepsilon)\,.
\end{align}
Here $\hat{x}(\varepsilon)$ is the leading correction to $\varepsilon=0$ value of $f_c$. 
It seems that the presence of the non-perturbative contribution would prevent us from finding the $\varepsilon$ order correction to the location of the YLE singularity. This is, however, not the case.   
Indeed, after expressing $\scv$ in terms of $x_c$ and $f_c$ 
\begin{align}
    \scv_c = \frac { t_c } {h_c^{1/\beta \delta}}  = \frac{ x_c} { f_c^{1/\beta \delta} } \,
\end{align}
we obtain 
\begin{align}
\label{Eq:zc_eps}
    \scv_c \approx  \scv_c^{\rm MF} \left[
    1+\frac{  27 \ln \left(\frac{3}{2}\right)-(N-1) (\ln 2-5 i \pi
   )}{9 (N+8)}\epsilon \right]\,
\end{align}
where the terms proportional to $\hat x(\varepsilon)$ {\it cancel}. At higher orders of the $\varepsilon$ expansion, this cancellation does not happen. This prevents us from extracting corrections beyond the linear order. For the special case of $N=1$,  Eq.~\eqref{Eq:zc_eps} was previously derived in Ref.~\cite{Rennecke:2022ohx}.  

For the absolute value of $\scv$, we get  
\begin{align}
  |\scv_c| \approx  |\scv_c^{\rm MF}| \left[1 +  \frac{ 27 \ln \left(\frac{3}{2}\right) -  (N-1)  \ln 2}{9
   (N+8)}\epsilon \right]\,.
  \end{align} 
  Note that, the slope of the function $|\scv_c(\varepsilon)|$  is negative for $N<1 + 27 \left(\frac{\ln 3}{\ln 2} -1 \right)\approx 16.8$. It demonstrates that, at least as long the slope of the $d$ dependence is concerned,  to reproduce the result of  $N\to \infty$ limit one has to consider rather large values of $N \gg 17$.  Note that the $\varepsilon$-expansion at any given $\varepsilon$ leads to a monotonic dependence of the location on $N$. As we demonstrate in Sect.~\ref{sec:ylesing}, for the physical point $d=3$ or $\varepsilon=1$, this dependence is non monotonic.

 Using Eq.~\eqref{Eq:Rchi} for the universal ratio $R_\chi$ leads to 
\begin{align}
  |\zeta_c| \approx  |\scv_c^{\rm MF}| \left[
  1-\frac{ 2 (N-1) \ln 2+27 \ln 3}{18 (N+8)} \epsilon 
  \right]\,.
  \end{align} 
In $N\to \infty$ limit, this results reproduces the leading order expansion near four dimension of Sect.~\ref{Sec:LargeN} in $d=4-\varepsilon$, see Eq.~\eqref{Eq:LargeN_near4}.

 \subsection{Behaviour near/at $d=2$}
 \label{sec:twod}
 For the Ising universality class, $N=1$, $d=2$ and $d\to 1^+$ are analytically treatable. We refer the reader to Refs.~\cite{Fonseca:2001dc,Xu:2022mmw} for $d=2$ and to Ref.~\cite{Rennecke:2022ohx} for $d\to 1^+$. In the latter reference, the two-dimensional results were also presented in the same normalization scheme as in this paper.

\begin{figure}
    \centering
    \includegraphics[width=0.49\linewidth]{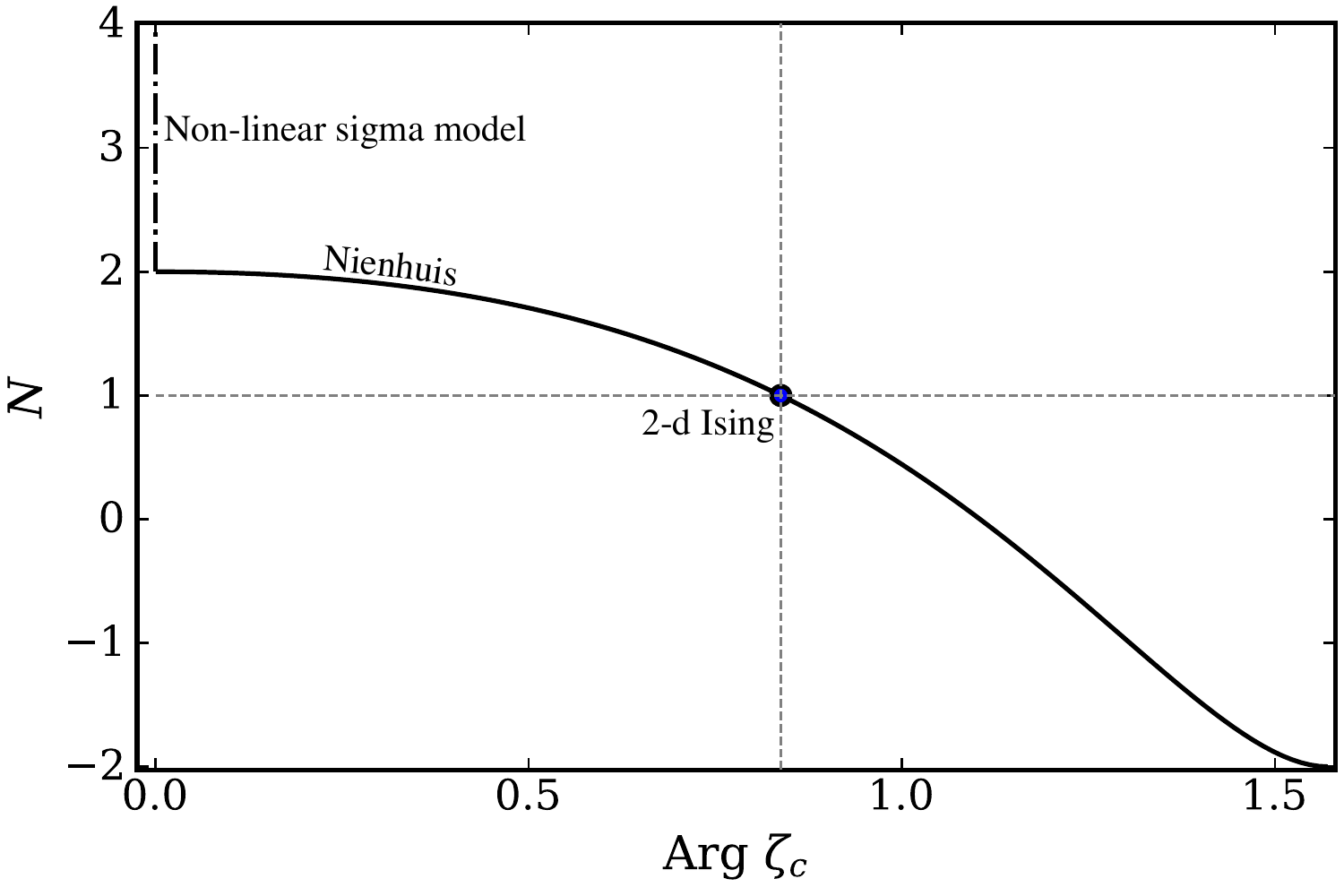}
    \includegraphics[width=0.49\linewidth]{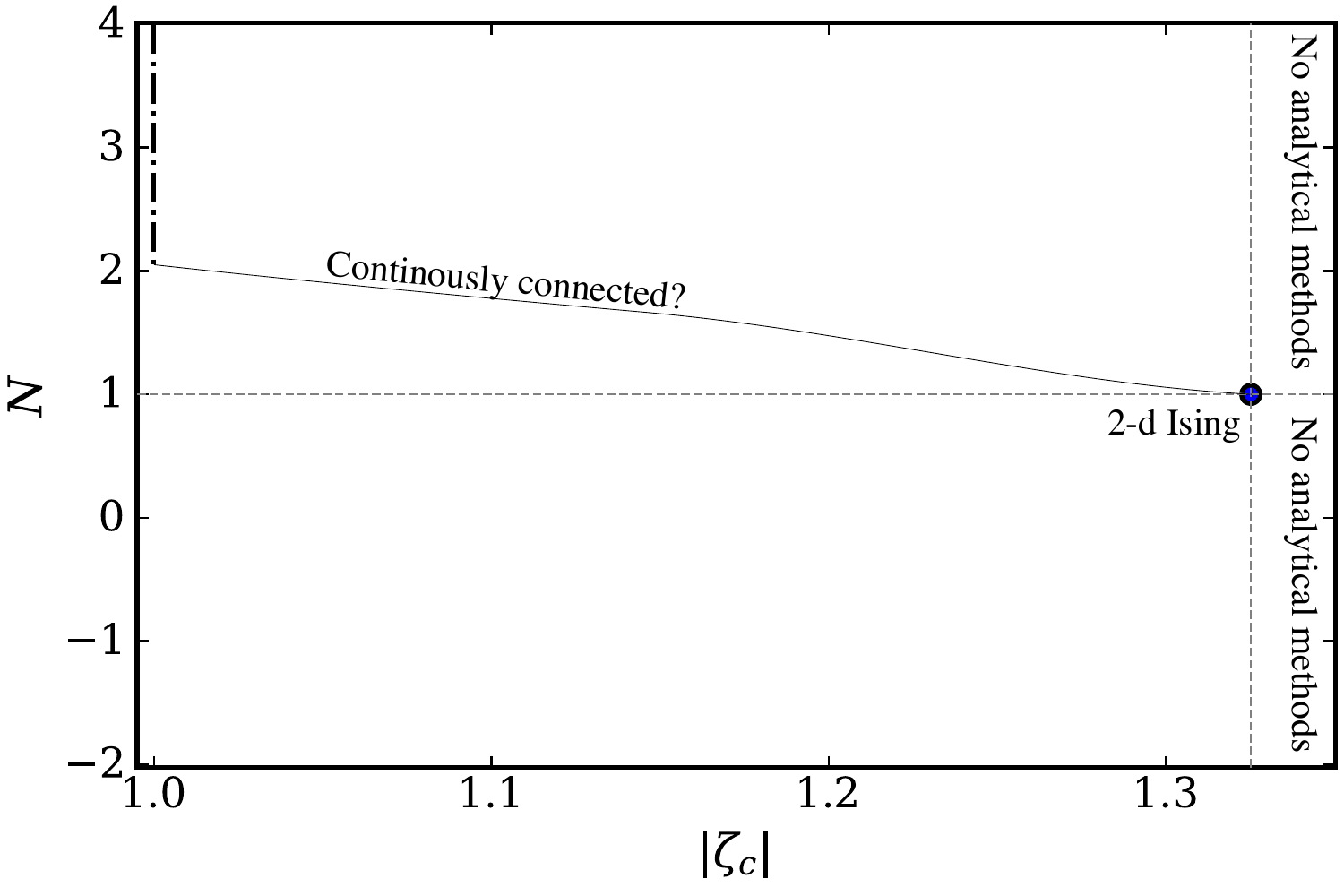}
    \caption{Argument and the absolute value of the Yang-Lee edge singularity at $d=2^+$; the Ising model $N=1$ is depicted by the  dots. 
    For $-2<N<2$, the argument of the YLE singularity is defined by the critical exponents of the underlying universality class, which are known exactly, see Ref.~\cite{PhysRevLett.49.1062}.  For $N>2$, the non-linear sigma model predicts the argument of the singularity. There are no exact results for the location of the singularity in $d=2$. For $N=1$, Ref.~\cite{Xu:2022mmw} provides the location with high precision. For $N>2$, we conjecture that $|\zeta_c|=1$. We expect to have a continuous connection between the start of the dash-dotted line and  $|\zeta_c|$ for $d=2$.  }
    \label{Fig:Arg_N}
\end{figure}

The case of $d=2$, $N\ne1$ deserves a special  attention.
\begin{itemize} 
\item  $N>2$: 
The perturbative analysis of the non-linear sigma model 
concludes that the theory near its lower critical dimension $d=2$ has a stable ultra-violet fixed point for $N>2$ with $N=2$ corresponding to Berezinskii–Kosterlitz–Thouless  phase transition~\cite{1971JETP...32..493B,1972JETP...34..610B,Kosterlitz_1973}. Complemented by the scaling relations this analysis also reveals the full set of critical exponents near two dimensions $d=2+\tilde \varepsilon$~\cite{PhysRevLett.36.691,PhysRevLett.45.499} 
\begin{align}
    \eta &= \frac{\tilde \varepsilon}{N-2} + {\cal O}(\tilde \varepsilon^0), \\ 
    \beta &= \frac{N-1}{2(N-2)} + {\cal O}(\tilde \varepsilon)\,.
\end{align}
Using Eq.~\eqref{Eq:etasc}, we can establish that to the leading order in $\varepsilon$
\begin{align}
    \Delta \approx \frac{2}{\tilde \varepsilon}
\end{align}
and thus $\Delta$ is $N$-independent (and coincides with the  $N\to \infty$ result). This fixes the argument of the Yang-Lee edge singularity to ${\rm Arg}\, \scv_c = \pi \tilde \varepsilon/4 \to 0$ at $d\to 2^+$.
By the analogy of the behaviour of the Ising model near its lower critical dimension and the result of $N\to \infty$ limit, see Eq.~\eqref{Eq:LargeN_near4}, we conjecture that  for any $N\ge 2$, $|\zeta_c| = |z_c| = 1$. We checked by direct analytic calculations that the location of the singularity in the non-linear sigma model for the Heisenberg model ($N=3$, see Ref.~\cite{Yang:1995qk} for the equation of state) follows our conjecture.     

\item $-2<N<2$:
For $-2<N<2$, the nonlinear sigma model does not have an ultraviolet fixed point (at least perturbatively). This, however,  does not exclude the presence of the Ising-like fixed point. We expect that the behaviour of the fractional $N$ smoothly interpolates between $N=1$ and $N=2^+$; for the location of the singularity, it means that $|\scv_c|$ changes from about 4 at $N=1$ to 1 at $N=2^-$. This plausible assumption is indirectly supported by the analytic results on the critical exponents, see Ref.~\cite{PhysRevLett.49.1062}~\footnote{See also FRG studies at fractional $N$ and $d=2$ in Refs.~\cite{Codello:2014yfa,PhysRevLett.110.141601,10.21468/SciPostPhys.10.6.134}.}, which 
behave smoothly as a function of $N$ in the range from $-2$ to $2^-$. Specifically, the argument of the location for the singularity 
\begin{align}
    {\rm Arg} \, \scv_c = \frac{\pi}{2\beta \delta}  =  \frac{4 \pi (2-v) v}{(v+1) (v+3)}
\end{align}
where $N= -2 \cos \frac{2\pi}{v} $ and $1<v<2$, 
smoothly traverses through $N=1$ (which corresponds to $v = 3/2$), where it accepts the two-dimensional Ising model's value ${\rm Arg} \  \scv_c = \frac{4\pi}{15}$. 
It then approaches 0 at $N=2^-$ ($v=2$), making a continuous connection to the result of the previous item, see Fig.~\ref{Fig:Arg_N}.

\item $N= -2 n$ where $n \in \mathbb{Z}^+$: In the absence of the magnetic field,  the theory with  negative even integer number of components is Gaussian for arbitrary $d$. Direct calculations show that $\gamma=1$ and $\eta=0$ and independent of $N$~\cite{PhysRevLett.30.544,Fisher:1973zzb}.  Using scaling relations, one thus finds that $\Delta = \frac{d+2}{4}$ and thus, in $d=2$, ${\rm Arg}\,  |\zeta_c| = \frac{\pi}{2}$. 

\end{itemize}

\section{Functional Renormalization Group}
 \label{Sec:FRG}

\subsection{Overview of the Functional Renormalization Group}
We briefly overview the Functional Renormalization Group (FRG) approach; for a thorough review, see Refs.~\cite{Wetterich:1992yh, Berges:2000ew, Delamotte:2003dw, Braun:2011pp, Kopietz:2010zz, Dupuis:2020fhh}.  
FRG is a specific field-theoretical implementation of Wilson's idea of integrating over momentum shells which is achieved by the inclusion of fluctuations ordered by momentum scales. Practically this is done through modifying the path integral measure by adding a mass-like term $\Delta S_k[\varphi]$ suppressing contributions of momentum modes with $p\lesssim k$. Under appropriate conditions on $\Delta S_k[\varphi]$, variation of the scale parameter $k$ will lead to an equation connecting the UV effective action at the  scale $\Lambda$,  $\Gamma_{k=\Lambda}[\phi] \approx S[\phi]$ to the full IR effective action at $k=0$, $\Gamma_{k=0}[\phi]=\Gamma[\phi]$. The expectation value (or the order parameter) $\phi$ is given by  $\phi(x)=\langle \varphi(x) \rangle$.

In the presence of $\Delta \, S_k[\phi]$, the partition function reads
\begin{eqnarray}
\mathcal{Z}_k[J]&=&\int\mathcal{D}\varphi \, e^{-\Delta S[\varphi]} \,\, e^{-S[\varphi]+J\varphi}
\end{eqnarray}
and thus becomes scale-dependent. Usually, the following choice is considered 
\begin{equation}
    \Delta S_k[\varphi] = \frac{1}{2}\int d^d x \int d^d y \,\, \sum_{i} {\varphi_i(x) R_k(x,y)\varphi_i(y)}. 
\end{equation}
In order the match the symmetry of the system, the mass-like regulator function $R_k(x,y)$ is chosen to be invariant under rotations (including the internal space rotations) and translations, i.e. $R_k(x,y)=R_k(x-y)$. Furthermore, in order to suppress modes with $p\lesssim k$ while leaving modes with $p \gtrsim k$ intact, the following must hold for the Fourier transform of the regulator:
\begin{eqnarray}
R_k(p) &\propto& \, k^2 \hspace{.5cm} \mbox{for} \hspace{.5cm}  p\ll k, \\
R_k(p) &\to& \, 0 \hspace{.5cm}  \mbox{for} \hspace{.5cm}  p\gg k.
\end{eqnarray}
$R_k(p)$ adds a mass of order $k^2$ to the low-energy modes, thereby suppressing their contributions to the path integral.

The  effective action, 
 $\Gamma_k[\phi]$ is obtained via a modified Legendre transform
\begin{eqnarray}
\label{Eq:gibbs}
\Gamma_k[\phi]&=-\mathrm{ln} \, \mathcal{Z}_k[J]+J \phi-\Delta S_k[\phi].
\end{eqnarray}
The functional $\Gamma_k[\phi]$ satisfies the Wetterich equation~\cite{Wetterich:1992yh,Morris:1993qb,Ellwanger:1993mw},
\begin{equation}
\label{Eq:flow}
    \partial_k\Gamma_k[\phi]=\frac{1}{2} \mbox{Tr} \Bigg\{ \partial_kR_{k}\bigg(\frac{\delta^2\Gamma_k[\phi]}{\delta \phi_i \delta \phi_j}+R_{k}\bigg)^{-1} \Bigg\}\,,
\end{equation}
also known as the flow equation. It prescribes the behaviour of $\Gamma_k$ between the classical tree-level action at an initial scale $k=\Lambda$ in the ultraviolet, $\Gamma_{k=\Lambda}=S$ -- an initial condition, and the desired full quantum action at $k=0$, $\Gamma_{k=0}=\Gamma$. The FRG equation provides a versatile realization of the Wilsonian RG and is as such well-suited to study critical physics. Both the scaling function and the critical exponents have been computed in great detail for O$(N)$ theories for real external fields with the FRG, see, e.g., Refs.~\cite{Berges:1995mw, Bohr:2000gp, Litim:2001dt, Bervillier:2007rc, Braun:2007td, Braun:2008sg, Benitez:2009xg, Stokic:2009uv, Litim:2010tt, Benitez:2011xx, Rancon_2013, Defenu:2014bea, Codello:2014yfa, Eichhorn:2016hdi, Litim:2016hlb, Juttner:2017cpr, Roscher:2018ucp, Yabunaka:2018mju, DePolsi:2020pjk}.

While the flow equation is exact, it defines an infinite tower of coupled partial differential equations for the effective action and its functional derivatives. With few exceptions (like the $O(N)$ model in the large-$N$ limit \cite{Tetradis:1995br}), truncation are therefore necessary in practice. There is often no obvious small parameter which can be used to define a systematic truncation scheme. Fortunately, this is not the case for critical physics, where the diverging correlation length facilitates a systematic expansion about vanishing momentum. Such a derivative expansion \cite{Morris:1994ie} has been shown to have a finite radius of convergence and a systematic error estimate is possible \cite{Balog:2019rrg, DePolsi:2020pjk}. We use the next-to-leading order of this expansion, i.e.\ first order in momentum-squared, in this work. This is elaborated in the next sections.

\subsection{First order derivative expansion}
\label{Sec:fde}
In order to solve the flow equation~\eqref{Eq:flow} numerically we will consider a constant field configuration around small momentum. The latter is only required to extract equations  
for the wave function renormalization. Specifically, we use the following anzatz for the scale-dependent effective action 
\begin{align}
    \label{Eq:truncation}
    \Gamma_k  = \int d^dx  \left[  \frac{1}{2} Z_k(\rho)\,  \partial^\mu \phi_a \partial_\mu \phi_a  +  \frac{1}{4} Y_k(\rho) \,  \partial^\mu \rho \partial_\mu \rho  + U_k(\rho) \right]\,, 
\end{align}
where $\rho=\frac{1}{2}\phi_a \phi^a$. The above expression contains all possible terms up to $\partial_\mu \partial^\mu$ in the derivative expansion. This approximation is appropriate for describing long-wavelength excitations in the critical region. 

Consider small deviations around the homogeneous field background, chosen to be non zero for the $i=1$ field component 
\begin{align}
    \phi_i(x) = \phi \delta_{i,1}  +  \varphi_i(x) 
\end{align}
then the wave-function renormalization for transverse and radial modes are 
\begin{align}
    Z^{\perp}_k(\rho) &= Z_k(\rho) = \lim_{p \to 0} \frac{\partial} {\partial p^2} \left. \frac{\delta \Gamma_k}{\delta ( \varphi_2(p) \varphi_2(-p))} \right|_{\varphi_a =0 }, \\
    Z^{\|}_k(\rho) &= Z_k(\rho) + \rho Y_k(\rho) = \lim_{p \to 0} \frac{\partial} {\partial p^2} \left. \frac{\delta \Gamma_k}{\delta ( \varphi_1(p) \varphi_1(-p))} \right|_{\varphi_a =0 }\,. 
\end{align}

\subsection{Flow equations for the next-to-leading order in the derivative expansion}

The flow equations for the potential can be obtained by substituting the truncation 
Eq.~\eqref{Eq:truncation} to Eq.~\eqref{Eq:flow} evaluated for a constant field configuration 
{
\begin{align}
\label{Eq:AverageAction}
    \partial_t U_k(\rho) &= \frac{1}{2} \int \bar d^dq\,   \partial_t R_k\left(q^2\right) \Big[G_k^\|+(N-1) G_k^\perp\Big]\,,
\end{align}
}
where 
\begin{align}
&G_k^{\perp} =  \frac{1}{Z_k^{\perp}(\phi)q^2 + U_k'(\phi)/\phi + R_k(q^2)} = \frac{1}{Z_k^{\perp}(\rho)q^2 + U_k'(\rho) + R_k(q^2)}, \\
&G_k^{\|} =  \frac{1}{Z_k^{\|}(\phi)q^2 + U_k''(\phi) + R_k(q^2)} = \frac{1}{Z_k^{\|}(\rho)q^2 + U_k'(\rho)+2\rho U_k''(\rho) + R_k(q^2)}.
\end{align}

For the integral measure, here and below we use  
\begin{align}
    \int \bar d^d q \equiv   \int  \frac{d^d q}{(2\pi)^d} \,. 
\end{align}

Introducing the tilde differential operator (see Ref.~\cite{Berges:2000ew})
\begin{align}
    \tilde \partial_t = \int d^d l \,   \partial_t R_k(l^2) \frac{\delta }{\delta R_k(l^2)}\,
\end{align}
allows us to write the equations in a succinct diagrammatic manner, 
\begin{align}
\nonumber
\partial_t \Gamma^{(2), \|}_k(p) = \frac{1}{2} \int \bar d^d q\,  \tilde \partial_t &\left(
{\hbox{\includegraphics{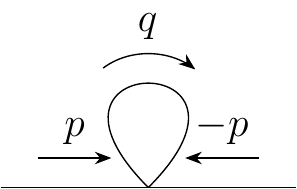}}}
-
\vcenter{\hbox{\includegraphics{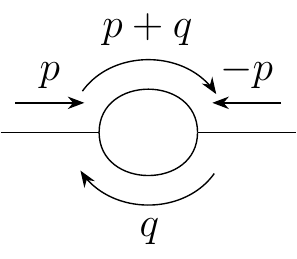}}}
\right. \\ 
&\quad \left. 
+
{\hbox{\includegraphics{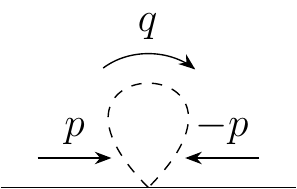}}}
 - 
 \vcenter{\hbox{\includegraphics{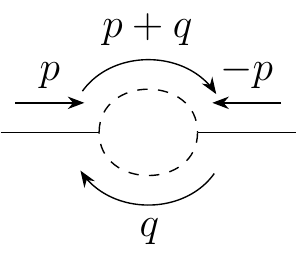}}}
\right), 
\end{align}
and 
\begin{align}
\nonumber
\partial_t \Gamma^{(2), \perp}_k(p) = \frac{1}{2} \int \bar d^d q \, \tilde \partial_t &\left(
{\hbox{\includegraphics{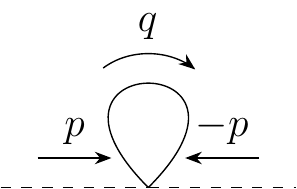}}}
-
\vcenter{\hbox{\includegraphics{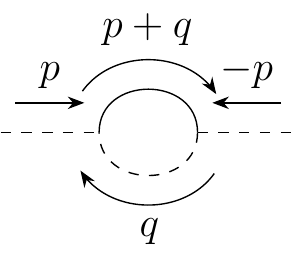}}}
\right. \\ 
&\quad \left. +
{\hbox{\includegraphics{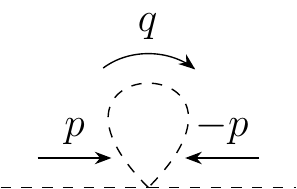}}}
-
\vcenter{\hbox{\includegraphics{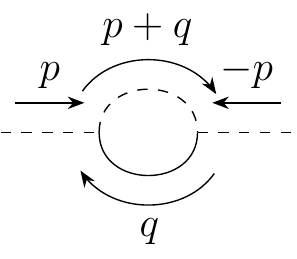}}}
\right)\,.
\end{align}
In the diagrams, the internal lines represent the scale-dependent Green functions for the transverse (solid line) and radial (dashed line) modes.
The vertices describing interaction relevant for the above flow equations are
\begingroup
\allowdisplaybreaks
\begin{align}
\Gamma^{(3)}_{1 1 1}(p,q) &= 
\left(
\vcenter{\hbox{\includegraphics{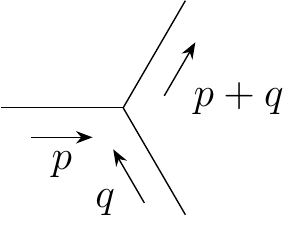}}}
\right)  \nonumber \\ &=  (p^2+q^2 + p \cdot q)  {Z_k^\|}'(\phi) + U_k^{(3)}(\phi),\\[8pt]
\Gamma^{(4)}_{1 1 1 1}(p,-p,q) &= 
\left(
\vcenter{\hbox{\includegraphics{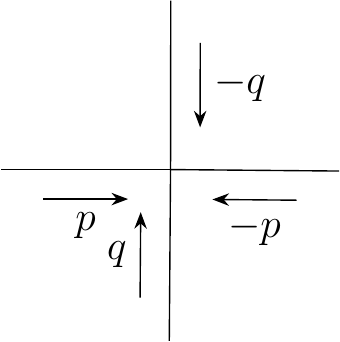}}}
\right)  \nonumber \\ &=(p^2+q^2) {Z_k^\|}''(\phi) + U_k^{(4)}(\phi),\\[8pt]
\Gamma^{(3)}_{1 i j}(p,q) &= 
\left(
\vcenter{\hbox{\includegraphics{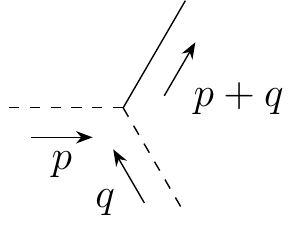}}}
 \right)  \nonumber \\ &= \left( (p+q)^2 \frac{ {Z_k^{\|}}(\phi) - {Z_k^{\perp}}(\phi) }{\phi}  - p\cdot q {Z_k^{\perp}}'(\phi)+ ( \frac{1}{\phi} U'_k(\phi))' \right) \delta_{ij},\\[8pt]
\Gamma^{(4)}_{i j}(p,-p,q) &= 
\left(
\vcenter{\hbox{\includegraphics{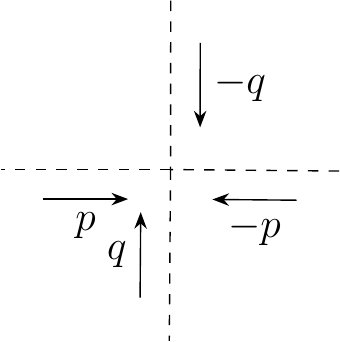}}}
 \right)  \nonumber \\ &= (p^2+q^2) \left[ 
2 \delta_{ij}\frac{Z_k^{\|}-Z_k^{\perp}}{\phi^2} + 
\frac{1}{\phi}{Z_k^{\perp}}'(\phi)
\right]+ \frac{1}{\phi} 
\frac{\partial }{\partial \phi} \left[ \frac{1}{\phi}  U'_k(\phi) \right]\,, \\[8pt]
\Gamma^{(4)}_{\|ij}(p,-p,q) &=  \left(
\vcenter{\hbox{\includegraphics{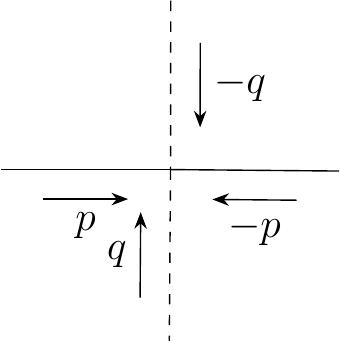}}}
 \right) \nonumber \\ &= \left(q^2 {Z_k^{\perp}}''(\phi) + p^2 \left[\frac{1}{\phi}{Z_k^{\|}}'(\phi)
 - 2 \frac{Z_k^{\|}-Z_k^{\perp}}{\phi^2}
\right]
+ (\frac{1}{\phi} 
 U'_k(\phi))''  \right) \delta_{ij}\,.
\end{align}
\endgroup
Using these vertices and applying the tilde derivative on the resulting expressions one can easily derive the flow equations for  $\Gamma^{(2), \|}_k(p)$ and $\Gamma^{(2), \perp}_k(p)$: 
\begin{align}
\notag 
\partial_t \Gamma^{(2), \|}_k(p) &=  \int \bar d^d q\, 
\partial_t R_k(q^2) \times \Bigg[
\\
&  
[G^\|(q)]^2
\left( 
- \frac12 \Gamma^{(4)}_{\|\|}(p,-p,q)
+ [\Gamma^{(3)}_{\|\|}(p,q)]^2 G^\|(p+q) 
\right) + 
\notag \\ &
(N-1)
[G^\perp(q)]^2
\left( 
- \frac12 \Gamma^{(4)}_{\|\perp}(p,-p,q)
+ [\Gamma^{(3)}_{\|\perp}(-(p+q),q)]^2 G^\perp(p+q) 
\right)
\Bigg], 
\end{align}
and 
\begin{align}
\notag 
\partial_t \Gamma^{(2), \perp}_k(p) &=  \int \bar d^d q\, 
\partial_t R_k(q^2) \times \Bigg[
\\
&  
[G^\|(q)]^2
\left( 
- \frac12 \Gamma^{(4)}_{\|\perp}(p,-p,q)
+  [\Gamma^{(3)}_{\|\perp}(p,-(p+q))]^2 G^\perp(p+q) 
\right) + 
\notag \\ &
[G^\perp(q)]^2
\left( 
- \frac12 (N-1) \Gamma^{(4)}_{\perp\perp}(p,-p,q)
+ [\Gamma^{(3)}_{\|\perp}(-(p+q),q)]^2 G^\|(p+q) 
\right)
\Bigg]\,.
\end{align}
Finally by taking the derivative of $\partial_t \Gamma^{(2), \|}_k(p)$ and $\partial_t \Gamma^{(2), \perp}_k(p)$  with respect to $p^2$ and evaluating at zero $p^2$ we arrive to the flow equations for the wave-function renormalizations: 
{
\begin{align}
    \label{Eq:Zperp}
    \notag 
     \partial_t Z_{\|}(\phi) &=
     \int \bar d^d q \partial_t R_k(q^2) \Bigg\{
        G_\|^2 \Big[ \gamma_\|^2 \big(G_\|' + 2 G_\|'' \frac{q^2}{d}\big) 
        + 2 \gamma_\| Z_\|'(\phi) \big(G_\| + 2 G_\|' \frac{q^2}{d}\big)  
         \notag \\ 
        & + (Z_\|'(\phi))^2 G_\| \frac{q^2}{d}
        - \frac12 Z''_\|(\phi) 
        \Big]\notag  \\&
        + (N-1) G^2_{\perp} \Big[ \gamma_\perp^2 \big(G_\perp' + 2 G_\perp'' \frac{q^2}{d}\big) 
        + 4 \gamma_\perp Z_\perp'(\phi)  G_\perp' \frac{q^2}{d}  
          + (Z_\perp'(\phi))^2 G_\perp \frac{q^2}{d}
        \notag \\ &
        + 2 \frac{Z_\|(\phi)-Z_\perp(\phi)}{\phi} \gamma_\perp G_\perp
        - \frac12 \left(\frac{1}{\phi}Z'_\|(\phi)
         - \frac{2}{\phi^2} (Z_\|-Z_\perp)
        \right)
        \Big] 
     \Bigg\}
     \\
     \label{Eq:Zpar}
      \partial_t Z^{\perp}_k(\phi) &= \int \bar d^d q \partial_t R_k(q^2) \Bigg\{G_\|^2 \Big[ \bar \gamma_\perp^2 \big(G_\perp' + 2 G_\perp'' \frac{q^2}{d}\big) 
        + 2 \bar \gamma_\perp Z_\perp'(\phi) \big(G_\perp + 2 G_\perp' \frac{q^2}{d}\big)  
         \notag \\ 
        & + (Z_\perp'(\phi))^2 G_\perp \frac{q^2}{d}
        - \frac12 Z''_\perp(\phi) 
        \Big]\notag \\
        & + 
        G^2_{\perp} \Big[ \bar \gamma_\perp^2 \big(G_\|' + 2 G_\|'' \frac{q^2}{d}\big) 
        +  4 \bar \gamma_\perp \left(\frac{Z_\|-Z_\perp}{\phi} -  Z_\perp'(\phi) \right)  G_\|' \frac{q^2}{d}  
         \notag  \\ &+ \left(2\frac{Z_\|-Z_\perp}{\phi}-Z_\perp'(\phi)\right)^2 G_\| \frac{q^2}{d}
        \notag \\ & 
        + 2 \frac{Z_\|-Z_\perp}{\phi} \bar \gamma_\perp 
        \big(G_\| + 2 G_\|' \frac{q^2}{d}\big)-\frac{Z_\|-Z_\perp}{\phi^2} - \frac12 (N-1) 
        \frac{1}{\phi}Z'_\perp(\phi)
        \Big] 
\end{align}
}
where we introduced short-hand notations for 
\begin{align}
    \gamma_\| &= q^2 Z_\|'(\phi) + U^{(3)}(\phi) \\
    \gamma_\perp &= q^2 Z_\perp'(\phi) + \frac{\partial}{\partial \phi} \left( \frac{1}{\phi}U' (\phi) \right)\,,
     \\
    \bar \gamma_\| &=  
    q^2 \frac{Z_\|-Z_\perp}{\phi}+ 
    U^{(3)}(\phi)
    \,.
    \\
    \bar \gamma_\perp &=  q^2 \frac{Z_\|-Z_\perp}{\phi}+\frac{\partial}{\partial \phi} \left( \frac{1}{\phi}U'(\phi) \right)\,.
\end{align}
as well as  to simplify the expression we denoted  
$G' = \frac{\partial G} {\partial q^2}$ and $G'' = \frac{\partial^2 G} {(\partial q^2)^2}$. 
In order to obtain Eqs.~\eqref{Eq:Zperp} and ~\eqref{Eq:Zpar} we applied the identity 
\begin{align}
    \int d^d q (p \cdot q)^2 f(q^2) 
     = \frac{p^2}{d }  \int d^d q q^2 f(q^2)\,.
\end{align}
and  used the expansion 
\begin{align}
    f\big((p+q)^2\big) =
    f(q^2) + (p^2 + 2 p\cdot q) f'(q^2) + 2 (p\cdot q)^2 f''(q^2) + {\cal O} (p^3)\,.
 \end{align}

In this paper, we consider the so-called strict derivative expansion. The logic behind this approximation is straightforward, see Ref.~\cite{DePolsi:2020pjk}. Below we rephrase it for the truncation of interest. 
At the $\partial^2$-order,  the momentum-dependent contributions of order $q^4$ to $\Gamma^{(4)}(p,q,q)$ are neglected; this justifies neglecting similar terms originating from the square of three-field vertex, $(\Gamma^{(3)}(p,q) )^2$. Dropping the corresponding terms we end up with   
\begin{align}
    \label{Eq:ZperpSDE}
    \notag 
     \partial_t Z_{\|}(\phi) &=
     \int \bar d^d q \partial_t R_k(q^2) \Bigg\{
        G_\|^2 \Big[ \gamma_\|^0 (2\gamma_\|-\gamma_\|^0)  \big(G_\|' + 2 G_\|'' \frac{q^2}{d}\big)
        + 2 \gamma_\|^0  Z_\|'(\phi) \big(G_\| + 2 G_\|' \frac{q^2}{d}\big)  
         \notag \\ 
        & 
        - \frac12 Z''_\|(\phi) 
        \Big]\notag  \\&
        + (N-1) G^2_{\perp} \Big[ \gamma_\perp^0 (2\gamma_\perp -\gamma_\perp^0) \big(G_\perp' + 2 G_\perp'' \frac{q^2}{d}\big) 
        + 4 \gamma^0_\perp Z_\perp'(\phi)  G_\perp' \frac{q^2}{d}  
        \notag \\ &
        + 2 \frac{Z_\|(\phi)-Z_\perp(\phi)}{\phi} \gamma^0_\perp G_\perp
        - \frac12 \left(\frac{1}{\phi}Z'_\|(\phi)
         - \frac{2}{\phi^2} (Z_\|-Z_\perp)
        \right)
        \Big] 
     \Bigg\},
     \\
     \label{Eq:ZparSDE}
      \partial_t Z^{\perp}_k(\phi) &= \int \bar d^d q \partial_t R_k(q^2) \Bigg\{G_\|^2 \Big[\bar \gamma^0_\perp (2\bar \gamma_\perp-\bar \gamma^0_\perp) \big(G_\perp' + 2 G_\perp'' \frac{q^2}{d}\big) 
        + 2 \bar \gamma^0_\perp Z_\perp'(\phi) \big({G_\perp} + 2 G_\perp' \frac{q^2}{d}\big)  
         \notag \\ 
        & 
        - \frac12 Z''_\perp(\phi) 
        \Big]\notag \\
        & + 
        G^2_{\perp} \Big[ \bar \gamma^0_\perp (2 \bar \gamma_\perp -\bar \gamma^0_\perp)\big(G_\|' + 2 G_\|'' \frac{q^2}{d}\big) 
        +  4 \bar \gamma^0_\perp \left(\frac{Z_\|-Z_\perp}{\phi} -  Z_\perp'(\phi) \right)  G_\|' \frac{q^2}{d}  
        \notag \\ & 
        + 2 \frac{Z_\|-Z_\perp}{\phi} \bar \gamma^0_\perp 
        \big(G_\| + 2 G_\|' \frac{q^2}{d}\big)-\frac{Z_\|-Z_\perp}{\phi^2} - \frac12 (N-1) 
        \frac{1}{\phi}Z'_\perp(\phi)
        \Big] \Bigg\}
\end{align}
where $\gamma_i^0 = \gamma_i(q=0)$. 

The final form of equations used for the flows of the expansion functions $U_k$, $Z_k^{\perp}$, and $Z_k^{||}$ in (\ref{Eq:AverageAction}) and (\ref{Eq:ZperpSDE})-(\ref{Eq:ZparSDE}), while written above in terms of the fields $\phi$, are re-expressed in terms of $\rho$ when probing the Wilson-Fisher point. In this form, the regularity of the flows at $\rho=0$ becomes apparent. With that said, the expressions in terms of $\phi$ are also necessary to probe the Yang-Lee edge singularity, as will be discussed.

\subsection{Regulator and wave function renormalization }
There are many different choices for the function $R_k(p)$. In this work, we consider Litim regulator~\cite{Litim:2001up} 
\begin{align}
  \label{Eq:Regulator}
    R_k(q^2) = a Z^\|_k (k^2 - q^2)
    \theta(k^2-q^2)\,.
\end{align}
Here $a$ is a parameter to be varied and optimized under the principle of minimal sensitivity~\cite{Canet:2002gs}, we come back to it in Sect.~\ref{Sec:minsen}. 
We note that we included the radial wave function renormalization at a given field background $\phi_0$,  $Z^\|_k = Z^\|_k(\phi=\phi_0)$, in the regulator. In this, we deviate from the conventional way when $Z_\perp(k)$ is introduced in the regulator. Our choice is shaped by the problem we are solving: at the YLE, it is expected that the transverse degrees of freedom decouple (the YLE is at a finite imaginary value of the magnetic field, see Fig.~\ref{fig:illustra}) while the radial mode is massless and thus dominant.  
It is convenient to explicitly separate $Z^\|_k$ from the field-dependent  $Z^\|_k(\phi)$. Moreover, we 
also normalize  $Z^\perp_k(\phi)$ by the same factor, that is 
\begin{align}
\label{Eq:norm}
    Z^{i}_{k}(\phi) = Z^{\|}_k z^{i}_k(\phi)  
\end{align}
where $i=\perp,\|$. From the definition of $Z^\|_k$, it follows that $z^{\|}_k(\phi_0) = 1$. 
At the same time, $z^{\perp}_k(\phi_0) \ne 1$ in general.

To simplify the equation, it is also convenient to introduce the ``renormalized'' field 
\begin{align}
\label{Eq:Ren}
    \phi_r = \sqrt{Z^\|_k} \phi 
\end{align}
This enables us to rewrite the Green functions in the following form 
\begin{align}
    G_\|^{-1}(q^2)
    = Z_k^\| [z_k^\|(\phi_r)  q^2  + U''(\phi_r) + a (k^2 - q^2) \theta(k^2-q^2)]
\end{align}
and 
\begin{align}
    G_\perp^{-1}(q^2)
    = Z_k^\| [ z_k^\perp(\phi_r) q^2  +  U'(\phi_r)/\phi_r + a (k^2 - q^2) \theta(k^2-q^2)]\,. 
\end{align}

At the fixed points, the anomalous dimension is related to the wave-function normalization through 
\begin{equation}
\partial_t Z^{\|}_k = - \eta_k   Z^{\|}_k\,.
\end{equation}
The anomalous dimension for the transverse component can be defined analogously, but it is not required for our needs.  

Note that 
\begin{align}
    Z_k^\perp(\phi)  = Z_k^\|(\phi) - \frac{\phi^2}{2} Y_k(\phi)  
\end{align}
and by analogy to Eq.~\eqref{Eq:norm}, it is  convenient to introduce 
\begin{align}
    y_k(\phi)   = \frac{Y_k(\phi)  } {Z^{\|}_k}\,.
\end{align}

\subsection{Truncation: notation and methodology} \label{Sec:truncation}
Equation \eqref{Eq:flow} is a differential (in $k$) and functional-differential (in the field space) equation. Its solution is not known. Without introducing a truncation scheme, this equation cannot be treated numerically. 

We thus perform a Taylor series expansion of the functions 
$z_\|^{(n)}(\phi)$, $y_\|^{(n)}(\phi)$ and $U(\phi)$ about a scale-dependent expansion point. In the vicinity of the Wilson-Fisher point, it is convenient to perform an expansion in terms of the ``renormalized'' field $\rho_r = \phi_r^2/2$, see Eq.~\eqref{Eq:Ren}. To simplify the notation, we omit the subscript and imply    $\rho_r\to\rho$ unless indicated explicitly.  
We have 
\begin{align}
    z^{\|}(\rho) &= 1 + \sum_{n=1}^{N_{Z}} \frac{1}{n!} z_n (\rho - \rho_{0,k})^n\,,   \\ 
    y(\rho) &= \sum_{n=0}^{N_{Y}} \frac{1}{n!} y_n (\rho - \rho_{0,k})^n\,,  \\
    U(\rho) &= \sum_{n=0}^{N_{U}} \frac{1}{n!} u_n (\rho - \rho_{0,k})^n\,.
\end{align}
In order to find the location of the YLE singularity, we also will perform the expansion in $\phi$, that is 
\begin{align}
    z^{\|}(\phi) &= 1 + \sum_{n=1}^{\underline{N_{Z}}} \frac{1}{n!} \underline{z}_n (\phi - \phi_{0,k})^n\,,   \\ 
    y(\phi) &= \sum_{n=0}^{\underline{N_{Y}}} \frac{1}{n!} \underline{y}_n (\phi - \phi_{0,k})^n\,,  \\
    U(\phi) &= \sum_{n=0}^{\underline{N_{U}}} \frac{1}{n!} \underline{u}_n (\phi - \phi_{0,k})^n\,.
\end{align}
We also omitted the subscript $k$, but emphasize that all parameters are running here.
In this paper, we require the truncation orders  to satisfy the requirement  $(\underline{N_U}, \underline{N_{Z_\|}},\underline{N_Y}) = 2 (N_U, N_{Z_\|}, N_Y)$. This guarantees that the truncations are consistent and one can perform switching between the variables without losing information about the corresponding function. 

The FRG flow equations for the coefficients can be readily derived starting from Eqs.~\eqref{Eq:AverageAction},  \eqref{Eq:ZperpSDE} and \eqref{Eq:ZparSDE}. 
For  renormalized quantities, we get a set of equations 
\begin{align}
\label{Eq:a}
    &\dot {u}_{n} -  u_{n+1} \dot { \rho}_0 =  \eta_\| ( u_{n+1} { \rho}_0 + n u_n) + 
     \frac{d^n}{d  \rho^n} \left( \partial_t U  \right)\Big|_{ \rho = \rho_0},\\
     \label{Eq:z}
    &\dot {z}^{i}_n - \eta_\| { z}^i_{n}
    - { z}^i_{n+1}
    \dot { \rho}_0 = 
     \eta_\| ( z^{i}_{n+1} { \rho}_0 + n  z^i_n) + 
   \frac{1}{Z_\|}\frac{d^n}{d  \rho^n} \left( \partial_t Z^i  \right)\Big|_{ \rho =  \rho_0}\,
\end{align}
where we introduced the polarization index $i=(\|,\perp)$. The coefficients $z_\perp^{n}$ are related to $y^{(n)}$ through $z_\perp^{n} =  z_\|^{n} - n y^{(n-1)} - \rho_0 y^{(n)}$.

In order to find the fixed point solutions, we also need to determine the flow equations for the dimensionless renormalized coefficients. They can be obtained by computing the expansion coefficient of dimensionless quantities, e.g. $U/k^d$ and $z^\|$, as functions of dimensionless $\tilde \rho = \rho_r / k^{d-2}$: 
\begin{align}
    \tilde u_n &= k^{n(d-2)-d} u_n \\
    \tilde z^{i}_n &= k^{n(d-2)} z^{i}_n \\
\end{align}
We have 
\begin{align}
\label{Eq:atilde}
    &\dot {\tilde u}_n - \tilde u_{n+1} \dot {\tilde \rho}_0 = 
    - d \tilde u_n + (d-2+\eta_\|) (\tilde u_{n+1} {\tilde \rho}_0 + n \tilde u_n) + 
    \frac{1}{k^d} \frac{d^n}{d \tilde \rho^n} \left( \partial_t U  \right)\Big|_{\tilde \rho = \tilde \rho_0},\\
    \label{Eq:ztilde}
    &\dot {\tilde z}^{i}_n - \eta_\| {\tilde z}^i_{n}
    - {\tilde z}^i_{n+1}
    \dot {\tilde \rho}_0 = 
     (d-2+\eta_\|) (\tilde z^{i}_{n+1} {\tilde \rho}_0 + n \tilde z^i_n) + \frac{1}{Z_\|}\frac{d^n}{d \tilde \rho^n} \left( \partial_t Z^i  \right)\Big|_{\tilde \rho = \tilde \rho_0}\,.
\end{align}
These equations can be rewritten in terms of the expansion coefficients of the series of $\phi$ without many modifications. It amounts to replacing $z^i_n, u_n$ and $\rho_0$ with  $\underline{z}^i_n, \underline{u}_n$ and $\phi_0$ and $\eta$ with $\eta/2$ in the right hand sides of  Eqs.~\eqref{Eq:a}, \eqref{Eq:z}, \eqref{Eq:atilde}, \eqref{Eq:ztilde}.

The equations above are to be supplemented by a choice of the expansion point. In this work we choose the expansion point $\phi_0$ by fixing the radial excitation mass, $m_R = \sqrt{U'(\rho)+2\rho\, U''(\rho)}$, to a constant, $\partial_t m_R = 0$. To approach the YLE fixed point and at the Wilson-Fisher point we set  $m_R$ to zero. In our work, the non-zero values of $m_R$ are only required to determine the critical exponent $\delta$ and most significantly the metric factor $B_c$.

The conventional expansion scheme, which is used in the majority of applications, follows the physical point defined by the minimum of the effective potential. Since the YLE singularity is located at the (imaginary) magnetic field $h_c$ where the radial mass at  vanishes, finding it requires numerically expensive fine-tuning with this scheme.

The advantage of our expansion scheme is that we can directly follow the flow of the edge singularity. Since the magnetic field enters as a linear term in the effective action, it is not renormalized in the flow equation \eqref{Eq:flow}. We can therefore simply read-off $h_c$ as the magnetic field which turns our expansion point $\phi_0$ into the physical point in the IR. The Lee-Yang theorem guarantees that we only have to resolve the effective action for purely real or purely imaginary fields in the symmetric phase, so full information in the complex plane is not required.

A downside of our expansion scheme is that numerical computations in the broken phase are more challenging. The reason is that in this case at any finite $k$ our expansion point $\phi_0$ lies on the real axis below the physical point. The FRG flow flattens the potential in this region since it has to be convex in the deep IR. This convexity-restoring flow is driven by near-singular propagators and therefore numerically challenging to resolve, see, e.g., Refs.~\cite{Litim:2006nn, Pawlowski:2014zaa}. However, as we have argued in Sect.~\ref{Sec:Scaling}, we do not need to compute in the broken phase.

\begin{figure}
    \centering
    \includegraphics[width=0.9\linewidth]{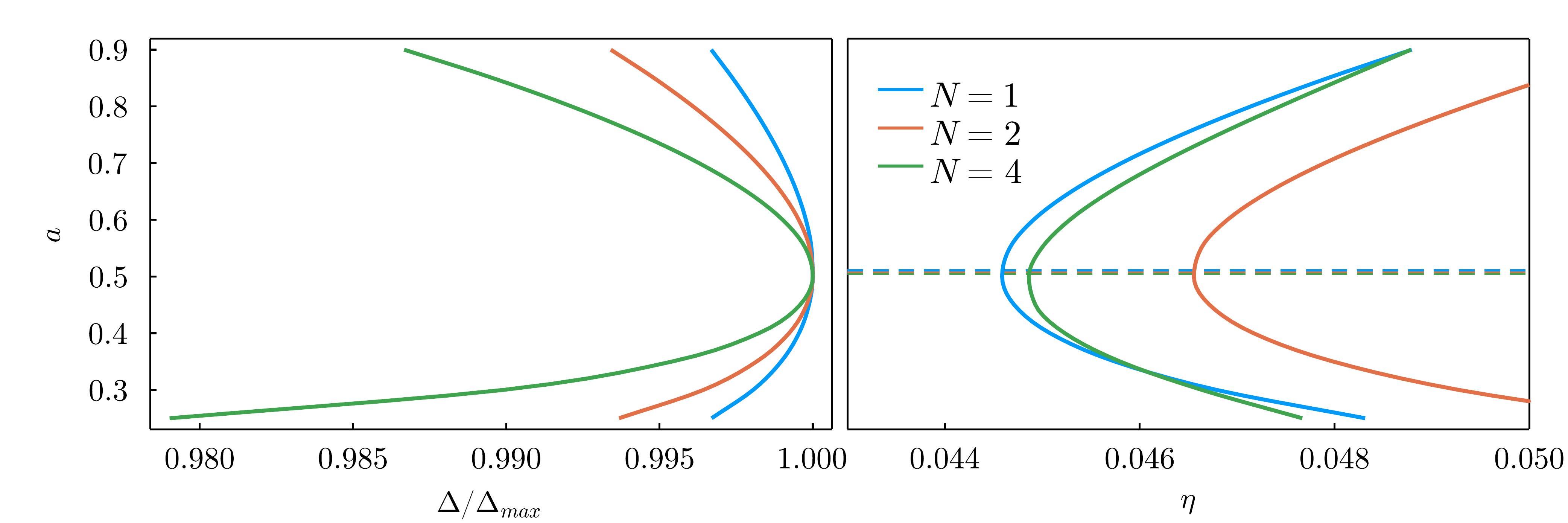}
    \caption{
    Regulator parameter dependence of the gap critical exponent $\Delta$ (normalized by its value at the maximum) and the anomalous dimension $\eta(a)$ for a set of $N$ at the Wilson-Fisher fixed point. The maxima of $\Delta$ define the minimal sensitivity point, $a_\Delta$. 
    The dashed lines in the right panel show $a_\Delta$. 
    The figure demonstrates that the location of the minima of $\eta(a)$ is fairly close to   $a_\Delta$.   
    } 
    \label{Fig:MSA}
\end{figure}

\subsection{O(N) fixed point and Minimal Sensitivity Analysis }
\label{Sec:minsen}
The Wilson-Fisher fixed point can be found by solving the set of algebraic equations 
$\dot{\tilde{u}}_n = \dot{\tilde{z}}^i_n=  \dot{\tilde{\rho}}_0 = 0$. 
There are several ways how we are going to use this solution. First, it defines the O(N) anomalous dimension; we use this critical exponent to apply the minimal sensitivity analysis via the regulator parameter $a$; see below. Second, we can use a slightly perturbed fixed point solution as the initial condition for the FRG evolution towards the IR to extract the metric factors, critical exponents, and finally the location of the YLE singularity.

\begin{figure}
    \centering
    \includegraphics[width=0.49\linewidth]{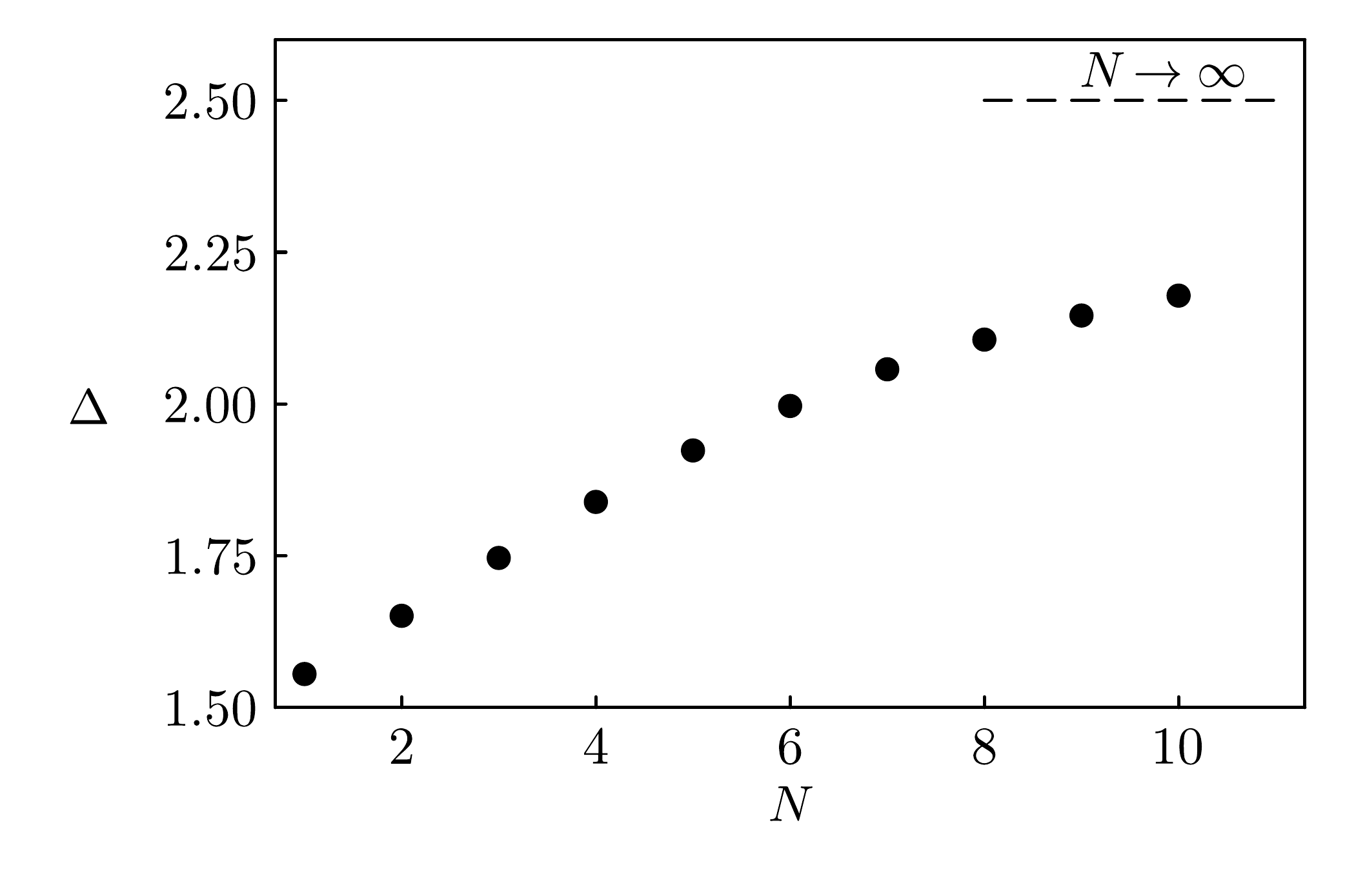}
    \includegraphics[width=0.49\linewidth]{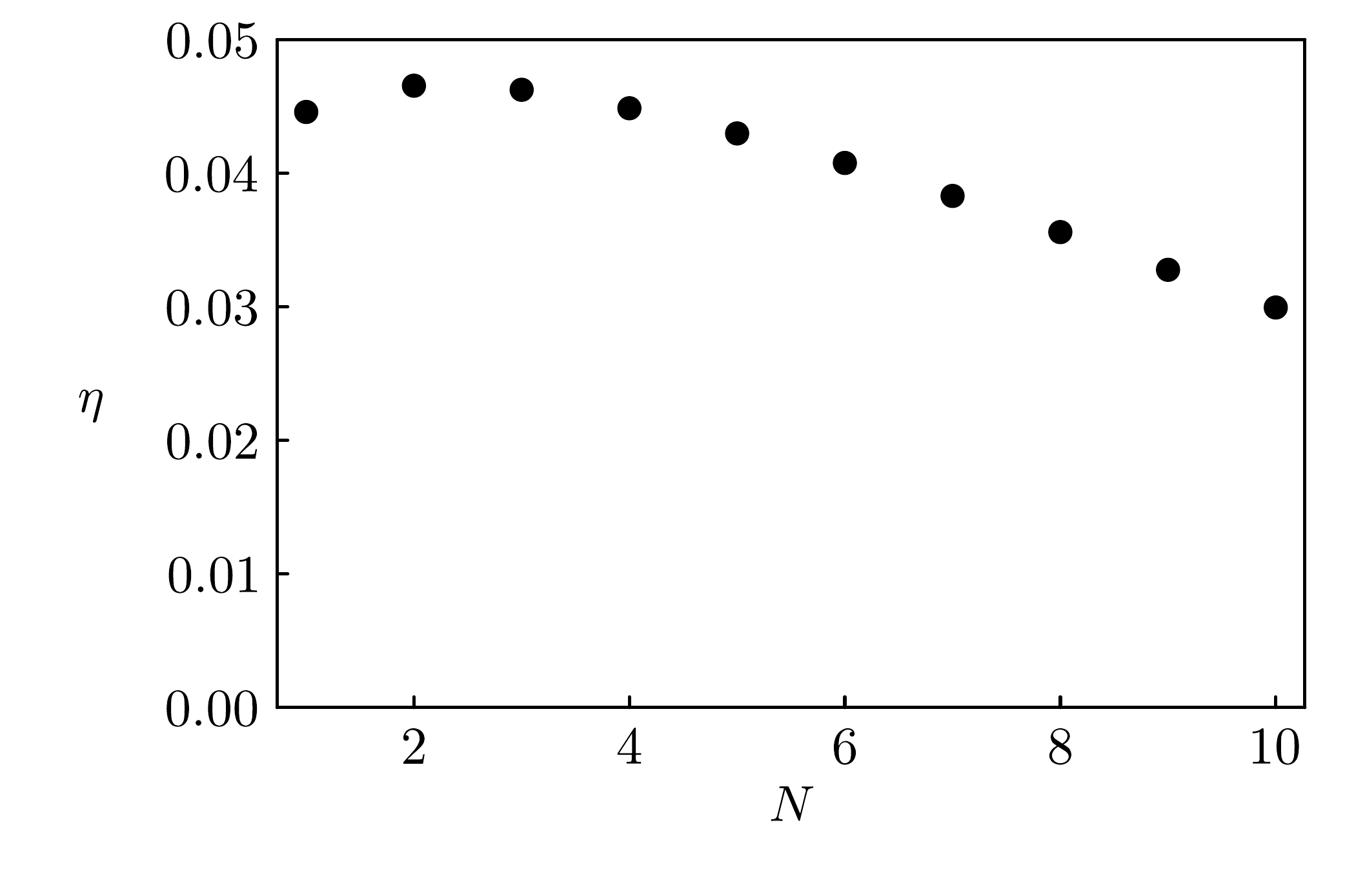}
    \caption{
    The results of the minimal sensitivity analysis at the Wilson-Fisher fixed point for the gap critical exponent $\Delta=\Delta(a_\Delta)$. The corresponding anomalous dimension is shown on the right panel $\eta=\eta(a_\Delta)$.
    }
    \label{Fig:MSA2}
\end{figure}

The critical exponent $\eta$ fully defines $\delta$ through the scaling relation  Eq.~\eqref{Eq:etasc}. 
Moreover, calculating the stability matrix at the fixed point solution allows one to find the critical exponent $\nu$ and thus the gap critical exponent through the relation 
$\Delta = \frac{\nu}{2} (d+2-\eta)$. 
 This motivates our strategy in defining the parameter $a$ as the extremum of the function $\Delta(a)$ where $a$ enters to the regulator through Eq.~\eqref{Eq:Regulator}. 
This fixes the regulator parameter $a=a_\Delta$ that we use for calculating the metric factors ($B_c$ and $C_+$),  critical exponents ($\delta$ and $\gamma$), and the location of the YLE singularity. By choosing the extremum as a function of $a$, it is guaranteed that among the family of regulators defined in Eq.~\eqref{Eq:Regulator}, we use the one where the regulator dependence, and hence the systematic error, is minimal at least for $\Delta$. 
An alternative approach would be to use minimal sensitivity analysis for all universal quantities, this, however, is not feasible for the location of the YLE singularity as it is defined through the (non-universal) metric factors and the critical exponents. 
 
We show the dependence of the gap critical exponent and the anomalous dimension on the regulator parameter $a$ in Fig.~\ref{Fig:MSA} for a few values of $N$. These calculations were performed using the strict derivative expansion and the truncation scheme $(N_U, N_{Z_\|}, N_Y) = (6,3,2)$. As can be seen in the figure, for $N$ of phenomenological interest, the location of the extrema of $\Delta(a)$ and  $\eta(a)$ are fairly close at this truncation. 

We obtained reasonable values of the anomalous dimension critical exponent displayed in Fig.~\ref{Fig:MSA2}. 
The non-monotonic dependence of $\eta$ on the number of components $N$ is expected from the $\varepsilon$-expansion.   

The minimal sensitivity analysis for different quantities does not necessarily result in the same regulator parameter $a$. In order to estimate the corresponding systematic uncertainty we also performed the analysis for the value of the magnetic field at the YLE singularity, see Table~\ref{Tab:areg}.

\subsection{Yang-Lee edge singularity fixed point solution} \label{Sec:YLEfp}
Near and at the YLE fixed point, due to explicit symmetry breaking $h\ne0$,  there is only one light degree of freedom -- the radial mode. Therefore we expect our result for the critical exponent $\eta_{\rm YLE}$ (or $\sigma_{\rm YLE}$)  to trivially reproduce those of the one component theory. This also serves as a cross-check on our calculations as it provides a powerful constraint on all our equations for $N\ne 1$.  

There is one subtle point related to how one approaches this fixed point. We remind the reader that at the Wilson-Fisher fixed point, the FRG equations are $k$-independent for properly scaled variables as we introduced in Sec.~\ref{Sec:truncation}.   The $k$ independence implies that the fixed point can be reached at a finite value of $k$.  To the contrary, when we start from the general equations for the multi-component field theory, one cannot expect $k$-independence for the YLE fixed point, as the complete separation of the transverse degrees of freedom is only possible at asymptotically small $k$. Thus, when finding the algebraic equation for the YLE  fixed point, one additionally has to take the limit $k \to 0$  for the terms that involve transverse degrees of freedom. 
In effect, this amounts to taking the limit of the dimensionless renormalized Goldstone mass to infinity. By computing this limit, we were able to show explicitly that the fixed point equations of our theory reduced to those of the single component theory, that is, the $N$-dependent terms originating from Eqs.~\eqref{Eq:Zperp} and    \eqref{Eq:AverageAction}  drop out.   

\subsection{Critical exponents and metric factors}
    \label{sec:crexp}
The scaling variable $\zeta$, defined in Eq.\ \eqref{Eq:zetaAmpl}, requires the determination of the critical exponents and metric factors. 
They can be found by performing calculations near the critical point. Thus appropriate perturbations of the initial conditions near the Wilson-Fisher fixed  point allow us to extract the required quantities. The numerical procedure, common for arbitrary $N$, coincides with that performed in Ref.~\cite{Rennecke:2022ohx}. 


\begin{figure}
    \centering
    \includegraphics[width=0.49\linewidth]{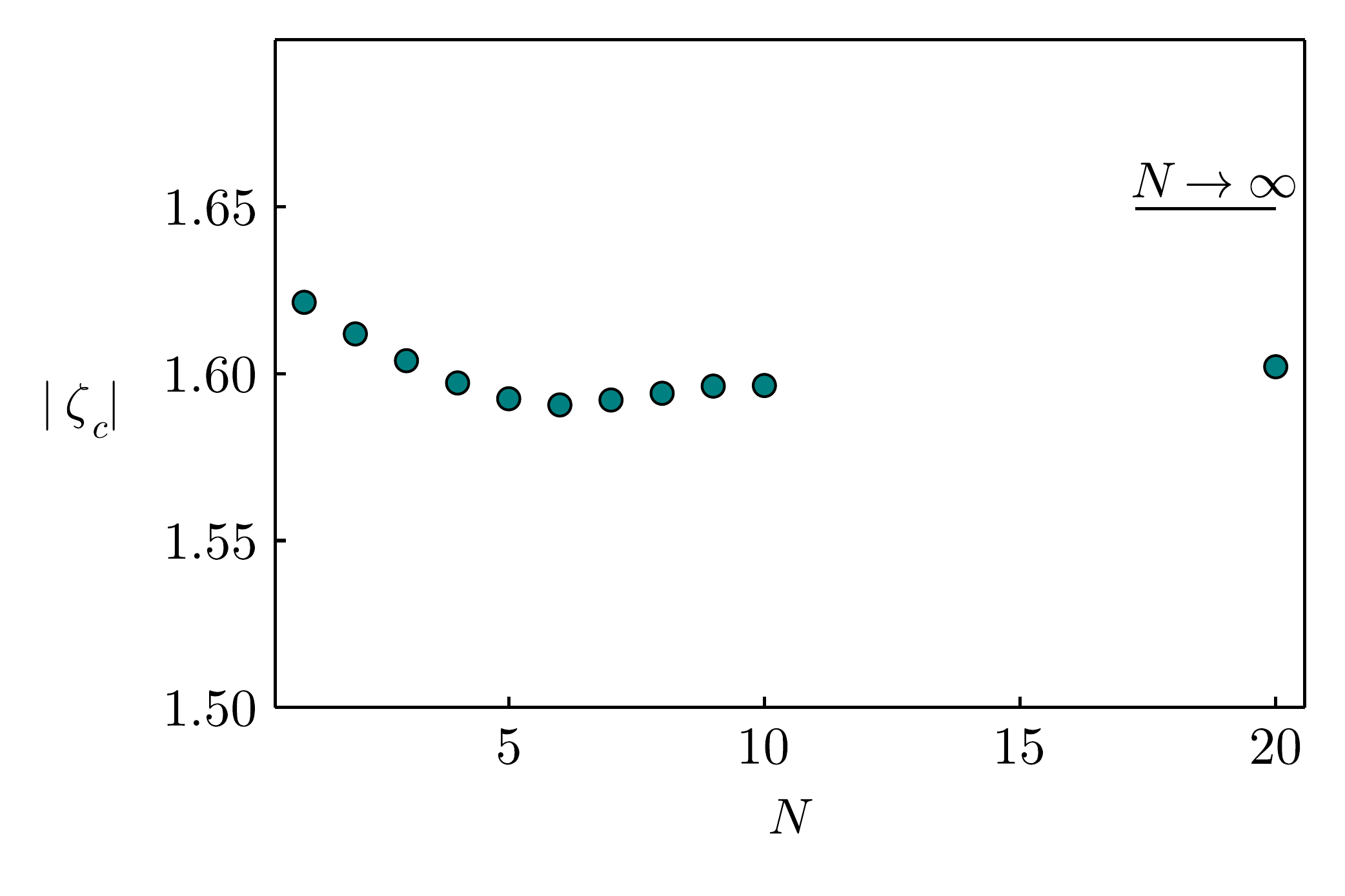}
    \caption{
    Location of the YLE singularity $|\zeta_c| = |z_c|/R_\chi^{1/\gamma}$ as a function of $N$. The estimated uncertainty is withing the marker size.   The infinite $N$ limit ($|\zeta_c| \approx 1.649$)  is approached from below. As was demonstrated in Ref.~\cite{Connelly:2020gwa}, this approach is parametrically slow.   
    }
    \label{Fig:YLE}
\end{figure}

\begin{table}
    \centering
    \begin{tabular}{|c|c|c|c|c|}
         \hline
         $N$& 1 & 2 & 3 & 4 
             \\
         \hline
         \hline
         $a_\Delta$& 0.5108 &  0.5069 & 0.5026 & 0.5044 
         \\ 
          \hline
         $a_\eta$& 0.5044 & 0.5075 & 0.5064 & 0.4906 
         \\ 
         \hline
          $a_{h}$& 0.6299
          & 0.5921
          & 0.5724
          & 0.5617
         \\ 
         \hline
    \end{tabular}
    \caption{The regulator parameter as determined by minimal sensitivity analysis applied to the gap critical exponent, $a_{\Delta}$, and  the anomalous dimension, $a_{\eta}$, at the Wilson-Fisher fixed point  as well as  to the value of the magnetic field at YLE singularity, $a_{h}$. The numbers are quoted to the fourth digit. 
    }
    \label{Tab:areg}
\end{table}

\subsection{Location of  Yang-Lee edge singularity}
\label{sec:ylesing}
To locate the YLE singularity, we compute in the symmetric phase  $t>0$ and at vanishing renormalized  mass  of the radial excitations. Practically we use small but nonzero values $m_R = 0+$. As we are interested in extracting the universal location, we have to consider rather small positive $t$ to minimize non-universal contributions. Similarly to Ref.~\cite{Rennecke:2022ohx} we perform the switch of the parametrization of the solution from $\rho$ to $\phi$ at some small but non-zero negative value of $\rho^{s}_0$. We checked that our results are in sensitive to the variation of the value of $\rho^{s}_0$.

Solving the flow numerically yields  the determination of $h_{\rm YLE} = U'(\phi_0)$ in the infrared.   Performing the mapping to $\zeta_c$ (see Ref.~\cite{Rennecke:2022ohx} for details) we obtained the results presented in Table~\ref{Tab:zetac} and illustrated in Fig.~\ref{Fig:YLE}. The error  was computed as follows. 
First, to estimate the error due to the truncation of the field dependence, $\Delta_{\rm tr}$, {we compare the results in the (5, 2, 1) and  (5, 3, 2) truncation schemes to the ones obtained for (6,3,2). We use the result of the highest truncation (6,3,2) for central points and maximal of the absolute values of the two differences
$|\zeta_c|^{(6,3,2)} - |\zeta_c|^{(5,3,2)}$ and 
$|\zeta_c|^{(6,3,2)} - |\zeta_c|^{(5,2,1)}$ for the error estimate due to truncation.}
 Second, to evaluate the uncertainty associated with the regulator dependence $\Delta_{\rm reg}$, we perform calculations of $|\zeta_c|$ at two values of $a$: $a_{\Delta}$ determined by the 
minimal sensitivity analysis applied to the critical exponent $\Delta$ at the Wilson-Fisher fixed point (see  Sect.~\ref{Sec:minsen}) and $a_{h}$ determined by the dependence of the magnetic field at the YLE singularity $h_{\rm YLE}$. The difference between the corresponding values of $|\zeta_c|$  determines $\Delta_{\rm reg}$. The numerical values for the regulators determined by both schemes are listed in Table~\ref{Tab:areg}. 
 Both our errors are measures for the convergence of our truncation within the next-to-leading order of the derivative expansion. A meaningful estimate for the systematic error of the derivative expansion itself requires us to go to next-to-next-to-leading order~\cite{Balog:2019rrg,DePolsi:2020pjk}~\footnote{Since the anomalous dimension is always zero in the leading order derivative expansion, it is not a suitable truncation to describe the YLE singularity and therefore also not suitable for a meaningful error estimate.}. While this is required for a precision calculation, this is beyond the scope of the present work.

\begin{table}
    \centering
    \begin{tabular}{|c|c|c|c|c|c|}
         \hline
         $N$& 1 & 2 & 3 & 4 & 5 
             \\
         \hline
         \hline
         $|\zeta_c|$& 1.621(4)(1) & 1.612(9)(0) & 1.604(7)(0)  & 1.597(3)(0)     & 1.5925(2)(1) 
         %
         \\ 
         \hline
    \end{tabular}
    \caption{The location of the YLE singularity, $|\zeta_c| = |z_c|/R_\chi^{1/\gamma}$  for 
     a representative number of components $N$. The numbers in the parentheses ($\Delta_{\rm tr}$), ($\Delta_{\rm reg}$) show  the truncation error and the error due to residual regulator dependence. The uncertainty quoted in the text corresponds to the maximum of $\Delta_{\rm tr}$ and $\Delta_{\rm reg}$. In all considered cases, $\Delta_{\rm tr}$ is the largest. 
     For the three-dimensional Ising universality class $N=1$,  the result of the current work is consistent with the previous study of Ref.~\cite{Rennecke:2022ohx}. }
    \label{Tab:zetac}
\end{table}

The next step is to perform the transformation from $|\zeta_c|$ to $|z_c|$. This step requires the determination of $R_\chi$. Our current set up does not allow us to compute this quantity as it necessitates solving the FRG equation in the broken phase which cannot be done using an expansion point defined by equation $\partial_t m_R = 0$ at finite $k$. Fortunately, high precision calculations of $R_\chi$ were performed recently in Ref.~\cite{DePolsi:2021cmi} for $N=2,3,4$ and $5$.  We will use these results together with the value of $R_\chi$ computed for $N=1$ in Ref.~\cite{Seide:1998ir}. Reference \cite{Seide:1998ir} does not provide systematic uncertainty on the value of $R_\chi$; we estimated it by comparing to earlier calculations of $R_\chi$ in the LPA' FRG of Ref.~\cite{Berges:1995mw}. We list the results in Table~\ref{Tab:rchi}, where to find $R_\chi^{1/\gamma}$  the value of $\gamma$ was taken from Ref.~\cite{DePolsi:2020pjk}. With this we can perform the transformation to $|z_c|$. The result is presented in Table~\ref{Tab:zc}. Within the systematic uncertainty the values are consistent to our previous calculations in LPA', see Ref.~\cite{Connelly:2020gwa}.   

Tables ~\ref{Tab:zetac} and ~\ref{Tab:zc} constitute the main results of this paper. 


\begin{table}
    \centering
\begin{tabular}{|c|c|c|c|c|c|}
 \hline
 $N$& 1 & 2 & 3 & 4 & 5\\
 \hline
  \hline
 $R_\chi^{1/\gamma}$& 1.497(22)  & 1.26(5)& 1.140(34) & 1.058(21)  & 0.974(26) \\
 \hline
 \end{tabular}
    \caption{The combination required to map $\zeta_c$ to $z_c$. Critical amplitude $R_\chi$ and critical exponent $\gamma$ are obtained from Ref.~\cite{Berges:1995mw,Seide:1998ir} and precision calculations of Refs.~\cite{DePolsi:2020pjk,DePolsi:2021cmi}.}
    \label{Tab:rchi}
\end{table}

\begin{table}
    \centering
\begin{tabular}{|c|c|c|c|c|c|}
 \hline
 $N$& 1 & 2 & 3 & 4 & 5 \\
 \hline
  \hline
 $|z_c|$ & 2.43(4) & 2.04(8) & 1.83(6)& 1.69(3) &  1.55(4)\\
 \hline
 \end{tabular}
    \caption{Location of the YLE singularity,   $z_c$, at different $N$. The uncertainty is dominated by the uncertainty  in determination of $R_\chi$. 
    }
    \label{Tab:zc}
\end{table}

\section{Conclusions}
\label{Sec:Concl}
Using the Functional Renormalization Group,  we extended our previous results, see Refs.~\cite{Connelly:2020gwa} and \cite{Rennecke:2022ohx}, for the universal location of the Yang-Lee edge singularity in the most important three-dimensional classic  $O(N)$ symmetric universality classes to the {(truncated)} next-to-leading order  in the derivative expansion. Furthermore, we used the prescription of the principle of minimal sensitivity to minimize the regulator dependence of our results. Our method is best suited for investigating the symmetric phase, and thus our main results are that for $|\zeta_c|=\frac{|z_c|}{R_{\chi}^{1/\gamma}}$. We used the high precision FRG calculations of $R_\chi$  and $\gamma$ from Refs.~\cite{DePolsi:2020pjk,DePolsi:2021cmi} for $N$=2, 3, 4, and 5 in order to find the location $|z_c|$.
For $N=1$,  $R_\chi$ was obtained from Refs.~\cite{Berges:1995mw,Seide:1998ir}. 
 See Tables \ref{Tab:zetac} and \ref{Tab:zc} for the summery of the results. To date, these are the best estimates for the location of the YLE singularity in classical O($N$) universality classes in three dimensions. 

\begin{figure}
    \centering
    \includegraphics[width=0.49\linewidth]{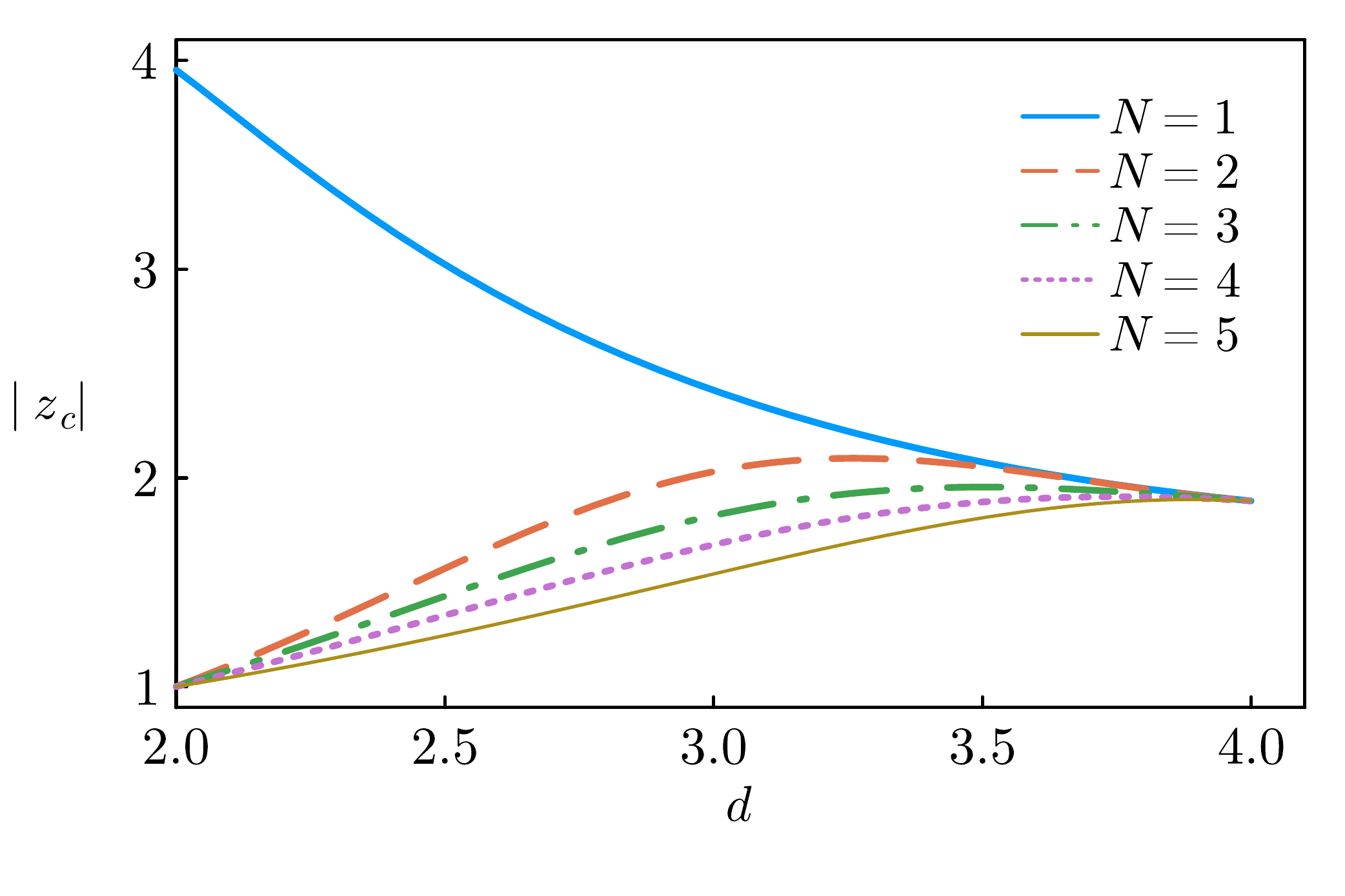}
    \caption{Four-parameter Pad\'e approximation for the dependence of the YLE location on the number of spatial dimension, see text for details.  The result for the location of the YLE singularity in two-dimensional Ising model is taken from Ref.~\cite{Xu:2022mmw}, see also Ref.~\cite{Rennecke:2022ohx} for reparametrization to $|z_c|$. 
    We expect that, at $d=2$, going to   fractional values of $N$ would fill in the gap from $N=1$ to $N=2$. 
    The line corresponding to $N=1$ reaches maximum at some value of $d$ in the range $1<d<2$ and then drops to 1 at the lower critical dimension $d=1$. 
    }
    \label{fig:yled}
\end{figure}

Combining the input from our FRG results with the {semi-exact} two-dimensional Ising model result of Ref.~\cite{Xu:2022mmw}, the epsilon expansion about $d=4$ (see Sect.~\ref{Sec:Epsilon}), and the behavior of $O(N>2$) systems near two dimensions (see Sect.~\ref{sec:twod}), we perform a Pad\`e approximation to capture  the dependence of the YLE singularity location, $|z_c|$, on the number of spatial dimensions, $d$, for $N=1-5$, see Fig.~\ref{fig:yled}. Qualitatively this result is similar to our earlier calculation of Ref.~\cite{Connelly:2020gwa}.

Complemented by recent ideas and methods of Refs.~\cite{Mukherjee:2019eou, Connelly:2020pno, Basar:2021hdf,Mukherjee:2021tyg,Dimopoulos:2021vrk,Nicotra:2021ijp,Basar:2021gyi,Bollweg:2022rps} our findings might help to establish the existence and potentially the location of the QCD critical point. 

\section{ Acknowledgement  }
We are  grateful
to  B.~Friman, T. Schaefer, and, especially, S.~Mukherjee for stimulating discussions leading to this work.  
We thank N. Wschebor for an illuminating discussion on strict derivative expansion.  
We acknowledge the computing resources provided on Henry2, a high-performance computing cluster operated by North Carolina State University, and support by  the U.S. Department of Energy, Office of Science,
Office of Nuclear Physics through the Contract~No.~DE-SC0020081.

\appendix

\section{On the non-perturbative nature of the $\epsilon$-expansion}
\label{sec:appendixnonp}
For the sake of simplicity, in this section we consider a special case of $N=1$. Our conclusions trivially extend to an arbitrary $N$.     

Consider the Widom equation of state (see e.g. Ref.~\cite{Guida:1996ep}): 
\begin{align}
    f(x) = &1 + x \notag \\ & + 
    \frac{1}{6} \epsilon  \left( - x \ln
   \left(\frac{27}{4}\right)+(x+3) \ln (x+3)-3 \ln (3)\right)
     \notag  \\ & + \frac{1}{648} \epsilon ^2 \left(9 (x+9) \ln ^2(x+3)+50 (x+3) \ln (x+3) + \ldots \right)
      \notag  \\ & + \epsilon^3 \left( 
      \frac{\ln^2(x+3) (675 + 246 x + 25 x^2)}{1944(x+3)} + \ldots \right) 
\end{align}
where in the two and  three loop contributions we explicitly displayed only a few principal terms.  
We also need the derivative 
\begin{align}
\label{Eq:dfN1}
f'(x) = & 1 
+\frac{1}{6} \epsilon  \left(\ln
   \left(\frac{4 (x+3)}{27}\right)+1\right)
   +\frac{1}{648} \epsilon ^2
   \left(\frac{4 (17 x+78) \ln (x+3)}{x+3}+\ldots \right)
    \notag \\ &+  
   \epsilon ^3  \left( \frac{(25 x (x+6)+63)  \ln ^2(x+3)}{1944 (x+3)^2}+\ldots \right)\,.
\end{align}
For the location of the singularity at one-loop order we then have
\begin{align}
    \beta \delta f(x_c)
    - x_c f'(x_c) = 
    \frac{x+3}{2} + 
    \frac{\epsilon}{12}   \left((x+9) \log (x+3) -x \left(2+\log
   \frac{27}{4}\right)-9 \log
   3 \right) =0 
\end{align}
with the approximate solution 
$x_c = -3 + \epsilon \left( \ln \frac{1}{\epsilon} + \ldots \right)$ where the ellipses include a constant term, nested logarithms and their ratios. 

Evaluating  $f'(x)$ at  $x_c = -3 + \epsilon  \ln \frac{1}{\epsilon}$ reveals the problem. In Eq.~\eqref{Eq:dfN1} let us separately  consider the two loop  
\begin{align}
  \left.   \frac{1}{648} \epsilon ^2
   \left(\frac{4 (17 x+78) \ln (x+3)}{x+3}+\ldots \right)\right|_{x\to -3 + \epsilon  \ln \frac{1}{\epsilon}} =-\frac{\epsilon }{6}-\frac{\epsilon  \log \left(\log \left(\frac{1}{\epsilon
   }\right)\right)}{6 \log (\epsilon )} + \ldots 
\end{align}
and the three-loop terms 
\begin{align}
  \left.   
  \epsilon ^3  \left( \frac{(25 x (x+6)+63)  \ln ^2(x+3)}{1944 (x+3)^2}+\ldots \right)
  \right|_{x\to -3 + \epsilon  \ln \frac{1}{\epsilon}} =
  -\frac{\epsilon }{12}-\frac{\epsilon  \log \left(\log \left(\frac{1}{\epsilon
   }\right)\right)}{6 \log (\epsilon )} + \ldots 
\end{align}
We see that despite the loop counting, both contributions are of order $\epsilon$ (that is the same as the one loop term)!  Higher order loop terms also contaminate $\epsilon^1$ order and similar terms involving nested logarithms. 
This necessitates all loop order resummation and thus brings us to the main conclusion of this appendix that the corrections to the location of YLE singularity are not perturbative in $\epsilon$. Note that this does not prevent us from extracting the leading order correction to $z_c$. This correction is of order of $\epsilon$; higher order contributions suffer from the non-perturbative contribution.

\newpage

\bibliography{YLEIsing}

\end{document}